\documentclass[10pt,twocolumn,groupedaddress,pra,aps,showpacs]{revtex4-1}
\usepackage{amsmath,mathrsfs,amsbsy,color,graphicx,bm,amsthm,amsfonts,dsfont,chngcntr}
\usepackage[english]{babel}
\usepackage{amssymb,lipsum,attachfile}
\usepackage{amsmath,epsfig,epstopdf}
\usepackage{amsfonts,dsfont,amsthm,bibunits}
\usepackage{hyperref}
\usepackage[normalem]{ulem}

\begin{document}
\newtheorem{theorem}{Theorem}
\newtheorem{proposition}[theorem]{Proposition}
\newtheorem{corollary}[theorem]{Corollary}
\newtheorem{open problem}[theorem]{Open Problem}
\newtheorem{Definition}{Definition}
\newtheorem{remark}{Remark}
\newtheorem{example}{Example}

\title{Topological qubits from valence bond solids}

\author{Dong-Sheng Wang}
\affiliation{Department of Physics and Astronomy, and Stewart Blusson Quantum Matter Institute, University of British Columbia, Vancouver, Canada}
\author{Ian Affleck}
\affiliation{Department of Physics and Astronomy, and Stewart Blusson Quantum Matter Institute, University of British Columbia, Vancouver, Canada}
\author{Robert Raussendorf}
\affiliation{Department of Physics and Astronomy, and Stewart Blusson Quantum Matter Institute, University of British Columbia, Vancouver, Canada}

\date{\today}


\begin{abstract}
    Topological qubits based on $SU(N)$-symmetric valence-bond solid models are constructed.
    A logical topological qubit is the ground subspace with two-fold degeneracy,
    which is due to the spontaneous breaking of a global parity symmetry.
    A logical $Z$-rotation by angle $\frac{2\pi}{N}$,
    for any integer $N>2$,
    is provided by a global twist operation,
    which is of topological nature
    and protected by the energy gap.
    A general concatenation scheme with standard quantum error-correction codes is also proposed,
    which can lead to better codes.
    Generic error-correction properties of symmetry-protected topological order are also demonstrated.
\end{abstract}

\maketitle

In recent years there has been significant interplay
between quantum computing
and topological states of quantum matter~\cite{ZCZ+15}.
Topological qubits have been proposed in various
systems~\cite{Kit03,IFI+02,MPM+05,AKT+08,DI12,BPI+14,OPH+16,Kap16},
and excitations, e.g., anyons,
support topologically protected gates by braiding~\cite{NSS+08}.
The valence-bond solids~\cite{AKLT87,AKLT88},
which are prototype models for matrix (and tensor) product states~\cite{PVW+07}
and allow symmetry-protected topological (SPT) order~\cite{CGW11,CGL+12,SPC11,DQ13a},
have also been exploited as
resources for quantum computing~\cite{RB01,WAR11,Miy11,WSR17}.

Fault-tolerant quantum computing and error-correction codes benefit from,
and also mostly rely on,
the stabilizer formalism and Pauli Hamiltonians~\cite{NC00}.
However, there are limitations~\cite{DKL+02,NO08,Ter15,BLP+16,ZCC11,EK09,AC16,DP17}:
e.g., the set of transversal qubit $Z$-rotations is restricted to be
of angles $2\pi/2^n$,
$n\in \mathbb{N}$~\cite{ZCC11,EK09,AC16}.
At the same time, more general coding theory has been developed
and non-stabilizer codes have been found~\cite{RHS+97,SSW07,CSS+09}.
In this work, we go beyond the stabilizer framework
and construct topological qubits and gates from
a class of $SU(N)$ valence-bond solids~\cite{GR07,KHK08,RSS+10,OT11,MUM+14} with 1D SPT order.
We find, for any integer $N>2$, there
exists a valence-bond solid qubit (VBSQ) such that the logical gate set
$\{\bar{X}, e^{i\frac{2\pi}{N} \bar{Z}}\}$ is transversal.
Larger sets of transversal gates can help reduce the circuit cost~\cite{ZCC11,EK09,AC16,FGK+15},
making fault-tolerant quantum computing more efficient.

\begin{figure}[b!]
\includegraphics[width=.25\textwidth]{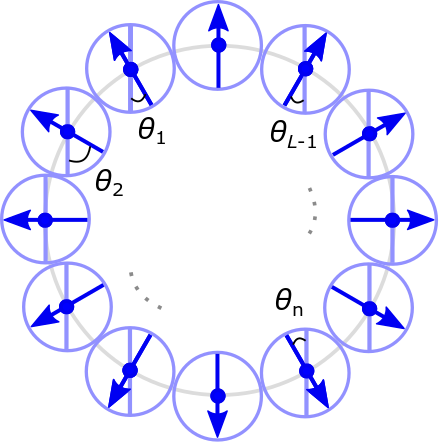}
\caption{Schematic diagram of a twist on a $SU(N)$ valence-bond solid qubit,
which executes the logical operation $e^{i\frac{2\pi}{N} \bar{Z}}$.
In general, a twist can be implemented by
a product of rotations around a fixed direction on every site,
with the rotation angles $\theta_n$ increasing smoothly from 0 to $2\pi\Omega$ along the system,
for $n=1,2,\dots,L$, $L$ the size of the system, and winding number $\Omega$.
The angle $\Phi$ of the logical rotation is
proportional to the winding number, $\Phi = 2\pi\, \Omega/N$.
}\label{fig:VBC}
\end{figure}

In this Letter, a VBSQ is based on the
degeneracy due to the spontaneous breaking
of a global parity symmetry,
while the logical space is also protected by a global $SU(N)$ symmetry.
We find both the broken and unbroken symmetries
provide transversal logical gates,
namely,
the logical $\bar{X}$ (bit flip) is the generator of the global parity symmetry,
while the logical $e^{i\frac{2\pi}{N} \bar{Z}}$
is provided by a global \emph{twist} operation (Fig.~\ref{fig:VBC}).
An appealing feature we show is that
the twist is topologically robust
and extracts the SPT order of ground states~\cite{CGW11,CGL+12,SPC11,DQ13a,WSR17}.
The gate $e^{i\frac{2\pi}{N} \bar{Z}}$ is implemented in a transversal and topologically stable fashion,
and it is furthermore outside the stabilizer formalism.
The existence of such gate implementations is a main finding of our work.

As a consequence of their SPT order and
distinct from topological stabilizer codes~\cite{Kit03,DKL+02,BK13},
VBSQs have a code distance that
grows linearly with the length of the code for bit flips (and a little more),
but is constant for generic phase flips,
which is due to the exponentially-decaying correlation functions.
Furthermore, VBSQs, viewed as a class of SPT codes,
and standard codes can be concatenated, improving error resilience.
That is, VBSQs provide error-protection at the hardware level,
afforded by an energy gap,
and error syndrome in addition.
By concatenation, bit flips are corrected at the hardware level,
while phase flips are corrected at the software level.
The situation encountered thus resembles classical hard drives,
which have a layer of physical error-correction
provided by the bulk magnetization of spins,
and a layer of software error-correction on top.

\begin{figure}[t!]
  \includegraphics[width=.45\textwidth]{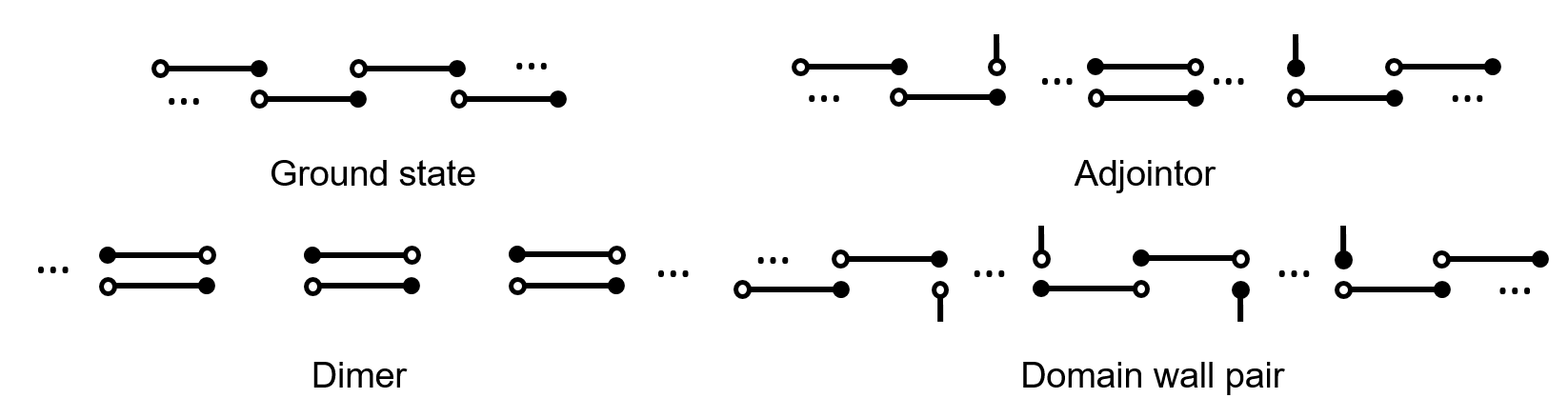}\vspace{-.5cm}
  \caption{Ground states and approximate excitations in the valence-bond picture.
  The on-site adjoint irrep is constructed from the projection
  from the product of a fundamental irrep (filled dot) and its conjugate (empty dot).
  Solid lines represent singlet (bond).
  (Up-left)
  A ground state that breaks $\mathbb{Z}_2^p$,
  and the other one is obtained by flipping each bond.
  (Up-right)
  A single adjointor excitation,
  whose size is confined by a linear potential,
  which is due to the dimerized region in between that has higher energy.
  (Down-left)
  A dimerized state,
  which preserves parity (reflection about a link)
  but breaks translation symmetry.
  (Down-right)
  A domain excitation with a pair of domain walls
  due to the periodic boundary condition,
  which roughly has the same energy as two adjointors.
  }\label{fig:vbs}
\end{figure}

We demonstrate our findings with the following main parts.
The code space is firstly defined,
and then transversal logical gates are studied.
Detailed analysis of the code properties then follows,
and finally, a concatenation framework is laid out
including a scheme of entangling gates.
The model we study is the translation-invariant $SU(N)$ valence-bond solid
\begin{equation}\label{eq:H}
  H_N:=\sum_{n=1}^L (J_1 h_n + J_2 h_n^2 + J_0 \mathds{1}_n):=\sum_{n=1}^L H_n
\end{equation}
on a ring of $L$ sites with periodic boundary
conditions (see Fig.~\ref{fig:VBC}) and
\begin{equation}\label{}
  h_n:=  \sum_{\alpha=1}^{N^2-1} T_n^\alpha \otimes  T^\alpha_{n+1},
\end{equation}
with $\{T^\alpha\}$ the generators of $SU(N)$
in the adjoint irrep for each site $n$~\cite{GR07,KHK08,RSS+10,OT11,MUM+14}.
For each $N>2$, and at $J_1=\frac{3N}{2}J_2=\frac{3}{N}J_0$
(w.l.o.g. let $J_1=1$),
the system has two degenerate ground states,
which span the code space of a VBSQ.
The two ground states,
denoted as $|\mathbf{L}\rangle$ and $|\mathbf{R}\rangle$,
are the spatial reflection of each other, namely,
they break a parity symmetry, denoted as $\mathbb{Z}_2^p$,
but preserve $SU(N)$ symmetry.
In the valence-bond picture (see Fig.~\ref{fig:vbs}),
the parity refers to the interchange of a fundamental irrep and its conjugate.
In terms of matrix product state we have
\begin{align}\label{eq:sptcode}
  |\mathbf{L}\rangle &=\vartheta\sum_{i_1=1}^{N^2-1}\cdots\sum_{i_L=1}^{N^2-1}
  \text{tr}(A^{i_1} A^{i_2}\cdots A^{i_L}) |i_1,\dots,i_L\rangle,
\end{align}
and $\{A^{i_n}\}$ is independent of the site label $n$,
which is then dropped,
and for each site
$  A^i=\sqrt{\frac{2}{N}} t^i,\; t^i=\lambda^i/2$
for the generalized Gell-Mann matrices $\{\lambda^i\}$,
while for $|\mathbf{R}\rangle$ the operators on each site
are $A^i=\sqrt{\frac{2}{N}} (t^i)^*$.
The normalization constant is
$\vartheta:=(\frac{N^2}{N^2-1})^{L/2}$
such that
the ground states are normalized and orthogonal
\begin{equation}\label{eq:overlap}
  \langle \mathbf{L}|\mathbf{L}\rangle=\langle \mathbf{R}|\mathbf{R}\rangle=1,\; \langle \mathbf{L}|\mathbf{R}\rangle=\left(\frac{1}{N-1}\right)^L\rightarrow 0.
\end{equation}
The model is frustration free;
hence its gap at zero temperature
is robust against local perturbation~\cite{MZ13}.
Some of its approximate excitations can also be obtained
(see Fig.~\ref{fig:vbs} and Supplementary Material~\cite{sm}, Sec.~I).

Now we define the VBSQs for each $N>2$.
In the large-$L$ limit, the operator
$P_\mathcal{C}:=|\mathbf{L}\rangle\langle \mathbf{L}|+|\mathbf{R}\rangle\langle \mathbf{R}|$
is the projection on the code space $\mathcal{C}$ since
the correction $\frac{1}{(N-1)^L}$ is exponentially suppressed.
Logical $\bar{X}$ operation
is the generator of the parity symmetry $\mathbb{Z}_2^p$.
In a certain on-site basis (see Supplemental Material~\cite{sm}, Sec.~I.B),
it can be expressed as a permutation operator $\Pi$ on each local site
\begin{equation}\label{eq:X}
\bar{X}|\mathbf{L}\rangle:= \Pi\otimes\Pi \cdots \Pi |\mathbf{L}\rangle = |\mathbf{R}\rangle,
\end{equation}
and
\begin{equation}\label{}
\Pi=\text{diag}(\sigma_x,\sigma_x,\dots,\sigma_x,\mathds{1}),
\end{equation}
for qubit Pauli operator
$\sigma_x=\left(\begin{smallmatrix}0&1\\1&0\end{smallmatrix}\right)$.
Notably, $\bar{X}$ is stable against single adjointor excitation
since there is no way for it to make a logical $\bar{X}$
without introducing a second one.
The situation is different if there are domains.
The size of a domain is not confined
as there is no binding potential between the two domain walls,
with one soliton (two empty dots)
and one antisoliton (two filled dots) (see Fig.~\ref{fig:vbs}),
although the size of domain walls is instead confined.
This means domain wall excitations can lead to logical bit flip error;
however,
the probability to induce $\bar{X}$
will be exponentially suppressed as the system size increases.
The system size should not be too big
in order to achieve a proper coding redundancy,
while also not too small to satisfy (\ref{eq:overlap}).

The $SU(N)$ symmetry provides the logical $\bar{Z}$-rotations by
a twist operator
\begin{equation}\label{eq:twi}
  U_\textsc{tw}:=\bigotimes_n e^{i\frac{2\pi}{L}n \mathcal{O}_n }:=\bigotimes_n U_n,
\end{equation}
and hermitian operator $\mathcal{O}_n$ on each site $n$ is
in Cartan subalgebra, i.e.,
it is diagonal,
and $e^{i 2\pi \mathcal{O}_n }=\mathds{1}$
(Supplemental Material~\cite{sm}, Sec.~II).
In the valence-bond picture,
the twist is equivalent to a unitary operator
\begin{equation}\label{eq:Vphase}
  V=\text{diag}(e^{i\ell}, 1,\dots, 1),\; \ell:= \frac{2\pi}{L},
\end{equation}
acting on each bond,
while $e^{i\ell}$ can be at any of the $N$
different positions (flavors) on the diagonal of $V$.
For any flavor we find
\begin{equation}\label{}
  \langle \mathbf{L}|U_\textsc{tw}|\mathbf{L}\rangle
  =e^{i2\pi/N},\;
  \langle \mathbf{R}|U_\textsc{tw}|\mathbf{R}\rangle
  =e^{-i2\pi/N},
\end{equation}
and $\langle \mathbf{L}|U_\textsc{tw}|\mathbf{R}\rangle=0$.
This provides the logical operator $e^{i\frac{2\pi}{N} \bar{Z}}$.
Actually, we see that the phases $e^{\pm i2\pi/N}$ are the SPT index
of the two ground states~\cite{DQ13a,WSR17,symm}.

The twist is a weak perturbation on the system.
The action of the twist is uniform on each bond,
while it is not the same on each physical site.
The disturbance of the twist to the local interaction terms $h_n$
is of $O(1/L^2)$ and negligible since
$[h_n, \mathcal{O}_{n+1}+\mathcal{O}_{n}]=0$,
and
$\langle \psi | [h_n, \mathcal{O}_{n+1}-\mathcal{O}_{n}]|\psi\rangle \in O(1/L)$,
$\forall |\psi\rangle\in \mathcal{C}$.
This means that the twist operator approximates the symmetry of the system
up to the $O(1/L)$ correction, which vanishes in the large-$L$ limit.

Furthermore, the twist operation is topological.
First, the phase gate $V$ (\ref{eq:Vphase}) can be slightly disturbed such that
the parameter $\ell_b$ for each bond $b$ is different and,
as long as $\sum_b \ell_b =2\pi$
and the second order $\ell_b^2$ is small,
the twist angle remains the same.
Second, the twist is also homotopic
since the same twist angle can be achieved on any long-enough continuous segment.
Third, if the set of $\ell_b$ sums to $2\pi f$ for
a fractional number $f\in (0,1)$,
the system will be excited and there will be an energy penalty.
This means the twist is protected by the gap.
If $f$ is slightly perturbed from 1 by $\delta f$,
the twist angle remains the same with corrections of the order $(\delta f)^2$.
Last but not least,
the twist is also stable against the single adjointor excitation.
When there are domains, however,
the phase accumulated from one ground state may cancel that from the other,
destroying the twist phase.

\begin{figure}
  \centering
  \includegraphics[width=.4\textwidth]{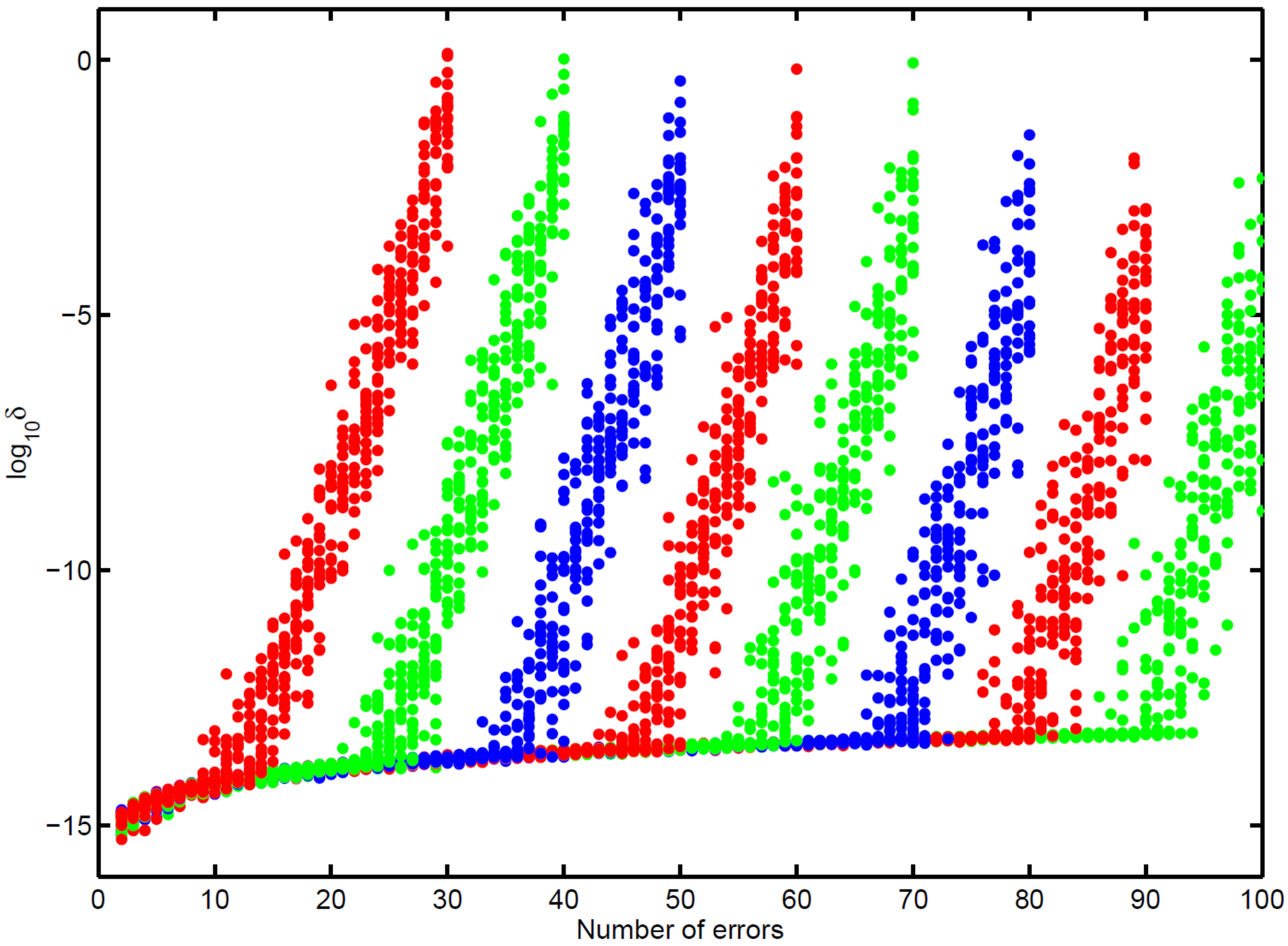}
  \caption{Simulation of the deviation $\log_{10}\delta$ of $\mathbf{U}(n)$ from identity
as the number of errors $n$ from the discrete error set $\mathcal{E}$~(\ref{eq:correctset})
 increases for different system sizes:
(from left to right) from $L=60$ to $200$ by an increase of amount $20$.
For a fixed $n$, we chose 20 random errors at $n$ random positions.
}\label{fig:Paulierror}
\end{figure}

Next we consider error-correction properties of VBSQ.
A correctable set of errors $\{E_i\}$ are defined such that
the error-correction condition
\begin{equation}\label{eq:KL}
  P_\mathcal{C} E_i^\dagger E_j P_\mathcal{C}= C_{ij} P_\mathcal{C}
\end{equation}
is satisfied,
and $\{C_{ij}\in \mathbb{C}\}$ form a hermitian matrix~\cite{KL97}.
Usually, a code can detect more errors that it can correct.
Different from error correction,
a detectable set of errors $\{E_i\}$ are defined such that
\begin{equation}\label{}
  P_\mathcal{C} E_i P_\mathcal{C}= e_i P_\mathcal{C}
\end{equation}
for $e_i\in \mathbb{C}$.
We study errors that are from independent local unitary errors.
In the following we mainly study three sets of errors:
errors from the set (\ref{eq:correctset}) below, $SU(N)$, and $SU(N^2-1)$,
respectively.
We find that the code distance is linear in $L$ for
the set (\ref{eq:correctset}),
while shows nontrivial error correction and detection
features for the other general errors.

First, we expect the VBSQ is robust against logical bit flip errors $\Pi$.
This is indeed the case (Supplemental Material~\cite{sm}, Sec.~III).
Consider the noncommutative error set
\begin{equation}\label{eq:correctset}
\mathcal{E}:=\{\Pi,\mathscr{P}^{jk}\}
\end{equation}
with $\mathscr{P}^{jk}$ as the adjoint rep of
the generalized Pauli operators $P^{jk}=X^jZ^k$
(Supplemental Material~\cite{sm}, Sec.~III).
Denote $E(n)$ as a product of $n$ errors acting on $n$ random positions
of the system with each from the set~(\ref{eq:correctset}),
and define the effective operator as
\begin{equation} \mathbf{U}(n):=\left(\begin{matrix}
   \langle \mathbf{L}|E(n)|\mathbf{L}\rangle &
   \langle \mathbf{L}|E(n)|\mathbf{R}\rangle \\
   \langle \mathbf{R}|E(n)|\mathbf{L}\rangle &
   \langle \mathbf{R}|E(n)|\mathbf{R}\rangle
\end{matrix}\right). \end{equation}
We find $\mathbf{U}(n)=(-\frac{1}{N^2-1})^n\mathds{1}$ holds till
a certain large enough $n$.
In particular, $\mathbf{U}(1)=-\frac{1}{N^2-1}\mathds{1}$,
$\mathbf{U}(2)=(\frac{1}{N^2-1})^2\mathds{1}$,
which shows that any single error from the set~(\ref{eq:correctset}) is correctable.
To determine its distance from the ideal identity gate,
we expand $\mathbf{U}(n)$ in terms of qubit Pauli matrices as
$\mathbf{U}(n)=\frac{1}{2}(s_0\mathds{1}+\vec{s}\cdot \vec{\sigma})$.
The numerical simulation in Fig.~\ref{fig:Paulierror} for the $SU(3)$ case
shows the value $\log_{10}\delta$
for $\delta:=\frac{|\vec{s}|}{|s_0|}$ for different system sizes.
We also find a similar behavior for the trace distance between
$\mathbf{U}(n)$ and identity.
We see that by an increase of $20$ for the system size,
the critical number of errors increases by $10$.
This shows that the critical number of errors is $L/2$,
which means the code distance for the error set~(\ref{eq:correctset}) is $L$,
saturating the classical Singleton bound~\cite{NC00}.

General errors lead to more complicated behavior.
For two local unitary errors $V, U \in SU(N)$
with a spatial distance $r$,
$\mathscr{V}$ and $\mathscr{U}$ as their adjoint rep,
we find
\begin{align}\label{}
\langle \mathbf{L}|\mathscr{V} \mathscr{U}|\mathbf{L}\rangle=\alpha+\beta,
\langle \mathbf{R}|\mathscr{V} \mathscr{U}|\mathbf{R}\rangle=\alpha+\beta^*,
\end{align}
and zero off-diagonal terms.
Here the parameters
$\alpha:=\frac{(N^2|u_0|^2-1)(N^2|v_0|^2-1)}{(N^2-1)^{2}}$,
$\beta:=(-\frac{1}{N^2-1})^{r+2} \beta_0$
for $\beta_0\in \mathbb{C}$,
and
$V=v_0 \mathds{1}+\sum_i v_i P_i$,
$U=u_0 \mathds{1}+\sum_i u_i P_i$,
expanded in the Pauli basis $\{P_i\}$ of the virtual space.
Note $\beta_0=0$ if $V=U$.
We see that the imaginary part $\beta$ decays exponentially
w.r.t. their spatial distance $r$.
The result above is consistent with
the exponentially-decaying correlation functions of the system,
and it means that arbitrary unitary errors in the symmetry
can cause a leakage together with a phase flip error,
and they are not exactly correctable.
However, the code will perform better for some cases,
e.g., when the errors are dilute which could be met by lowering the temperature.
This is confirmed by our simulations (Supplemental Material~\cite{sm}, Sec.~III.B).
Finally,
we expect on-site random unitary errors from $SU(N^2-1)$
are not correctable
since they destroy the SPT order of the system.
Indeed this is the case, and fortunately,
we find they are detectable (Supplemental Material~\cite{sm}, Sec.~III.B).

Error syndrome can be determined by the energy of local terms $H_n$.
For instance, after a Pauli error $\mathscr{P}$ on site $n$,
the energy of local terms $H_n$ and $H_{n-1}$ each
increases to $\frac{N^3(N^2+1)}{3(N^2-1)^2}$ (Supplemental Material~\cite{sm}, Sec.~III.C).
Recall that it is zero on ground states.
The total energy penalty is about twice
the energy of single adjointor excitation.
This means a Pauli error requires higher energy to occur
than an adjointor,
and Pauli errors can be suppressed by avoiding two or more adjointor excitations.
A general unitary error on site $n$ can also be detected
from the syndrome of $H_{n-1}$ and $H_n$.
Furthermore, if only one term shows syndrome,
this implies there is effectively only one adjointor,
given perfect energy check.
This causes no problem for error correction
since single adjointor cannot cause logical errors.
By cooling the system to a low temperature,
excitations will be suppressed and the Hamiltonian itself
provides protection of the code.

\begin{figure}[t]
  \centering
  \includegraphics[width=.4\textwidth]{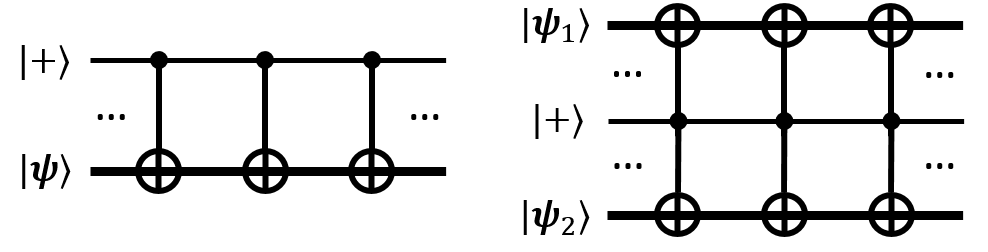}
  \caption{
  The measurement of $\bar{X}$ (Left) and $\bar{X}\bar{X}$ (Right).
  Here each controlled symbol is for the operator $\Pi$,
  and each controlled-$\Pi$ represent the transversal controlled-$\Pi$ between a particle in VBSQ
  and the five-qubit code (at its logical state $|+\rangle$).
  The particles in VBSQ are acted upon sequentially.
  }\label{fig:entangle}
\end{figure}

As a final part of this work,
we find the code performance of VBSQs can be improved
by the concatenation with an outer code, e.g.,
a certain stabilizer code.
A necessary ingredient of concatenation is
state preparation and coupling of many VBSQs.
We find there exists a measurement-based scheme~\cite{Ter15}
when $N=4k$, with positive integers $k\in \mathbb{N}$,
for which the logical $\bar{Z}$ gate can be realized by the twist
(Supplemental Material~\cite{sm}, Sec.~II).
Measurements of a logical $\bar{Z}$ or $\bar{X}$ will
prepare their eigenstates $|\mathbf{L}/\mathbf{R}\rangle$ or
$|\mathbf{U}/\mathbf{D}\rangle=
\frac{1}{\sqrt{2}}(|\mathbf{L}\rangle \pm |\mathbf{R}\rangle)$,
respectively.
Using the five-qubit code~\cite{NC00},
which has the parity property that
the Hamming weight of its logical $|0\rangle$ ($|1\rangle$) is even (odd),
a VBSQ at state $|\mathbf{L}\rangle$ can be prepared at state
$\frac{1}{\sqrt{2}}(|\mathbf{L}\rangle+(-1)^x|\mathbf{R}\rangle)$
by the transversal coupling between the five-qubit code
and each particle in VBSQ sequentially,
and measuring the logical $X$ of the five-qubit code
(see Fig.~\ref{fig:entangle}) at the end,
for $x$ as the parity extracted from the measurement outcomes.
This also works for other stabilizer codes that have the even-odd parity property,
such as Steane code and Shor code.
Error correction on the ancillary stabilizer code
can be performed during the process to enhance fault tolerance.
Logical operators $\bar{X}\bar{X}$, $\bar{Z}\bar{Z}$ etc for two VBSQs
can be measured in a similar fashion,
which can then realize logical entangling gates such as
the \textsc{cnot} and controlled-phase gates.

Given the robustness against the $\bar{X}$ error,
the repetition code is the most efficient choice
to deal with a $\bar{Z}$ error on a VBSQ.
For a series of VBSQ rings in parallel,
enforcing the stabilizers $\bar{X}_i\bar{X}_{i+1}$ for all nearest-neighbor pairs of rings
prepares the repetition code with codeword
\begin{equation}\label{}
  |0/1_\textsc{l}\rangle:=\frac{1}{\sqrt{2}}(|\mathbf{U}\mathbf{U}\mathbf{U}\cdots \rangle
\pm |\mathbf{D}\mathbf{D}\mathbf{D}\cdots \rangle),
\end{equation}
and the logical operator $X_\textsc{l}:=\bar{X}$ for $\bar{X}$ on any ring
and logical $Z_\textsc{l}:=\bar{Z}\bar{Z}\bar{Z}\cdots$.
Notably, our code can correct more $\bar{Z}$ errors.
When the parity check on the software level indicates a $\bar{Z}$ error,
it can then be located with the help of error detection of individual rings.
As a result, the minimal number of rings is two instead of three
for the repetition code,
and in general, it can correct up to $\kappa-1$,
instead of $\frac{1}{2}(\kappa-1)$,
phase flip errors for $\kappa$ rings.
More generally, concatenation with stabilizer codes
can be employed to safely avoid $\bar{X}$ error on a single VBSQ ring.
The error detection of single ring will also improve the code performance.
For transversal gates,
VBSQs for $N=2^{k+1}$ concatenated with stabilizer codes
allow transversal implementation of $Z$-rotation of angle $\frac{\pi}{2^k}$.
For instance, there is a transversal phase gate for $N=8$ and a $T$ gate for $N=16$,
which are constrained by the structure of the Clifford hierarchy
(also see Supplemental Material~\cite{sm}, Sec.~IV).

In summary, we have proposed a construction of valence-bond solid qubits
and concatenation schemes with stabilizer codes.
We notice a distinction for even and odd $N$,
while a scheme for odd $N$ cases remains to be discovered.
Our work demonstrates fundamental features of
valence-bond solids and symmetry-protected topological order
for quantum error-correction and quantum memory,
and can also be generalized to other valence-bond solids or crystals
for various symmetries or spatial dimensions.

This work is funded by NSERC and Cifar.
D.-S. W. acknowledges Y.-J. Wang for many suggestions,
N. Delfosse, L. Jiang, and
Z.-C. Gu for valuable discussions.

\bibliography{ext}{}

\begin{thebibliography}{56}
\bibitem [{sym()}]{symm}%
  \BibitemOpen
  \href@noop {} {}\bibinfo {note} {Note the symmetry used for SPT phase
  classification is SU($N$)$/Z_N$.}\BibitemShut {Stop}%
\end{thebibliography}
\bibliographystyle{apsrev4-1}

\end{document}


%

\title{Supplementary Material}

\maketitle

\newtheorem{theorem}{Theorem}
\newtheorem{proposition}[theorem]{Proposition}
\newtheorem{corollary}[theorem]{Corollary}
\newtheorem{open problem}[theorem]{Open Problem}
\newtheorem{Definition}{Definition}
\newtheorem{remark}{Remark}
\newtheorem{example}{Example}

\setcounter{figure}{0}
\setcounter{table}{0}
\setcounter{section}{0}
\setcounter{page}{1}
\renewcommand{\bibnumfmt}[1]{[R#1]}
\renewcommand{\citenumfont}[1]{R#1}

\renewcommand{\theequation}{SA\arabic{equation}}
\renewcommand{\thefigure}{SA\arabic{figure}}
\renewcommand{\thetable}{SA\arabic{table}}

\section{The system}
\label{sec:system}

In this section, we present details of the properties of the system,
including energy of states, correlation functions, excitations, and the parity symmetry.

\subsection{Energy of states}
\label{sec:gsenergy}

With the set of generalized Gell-Mann matrices $\{t^m\}$,
we define the following matrix
\begin{equation}
  \mathcal{M}:=\frac{2}{N}\sum_m t^m\otimes t^{m*},
\end{equation}
which is often called a ``transfer matrix.''
Its eigenvectors are easy to find to be
\begin{align}\label{eq:basiss}
  |v_0\rangle= & \frac{1}{\sqrt{N}}\sum_i |ii\rangle:=|\omega\rangle, \\
  |v_a\rangle= & \sqrt{2}\sum_{ij}t^a_{ij}|ij\rangle=\sqrt{2N} (t^a\otimes \mathds{1})|\omega\rangle,
\end{align}
with $\langle v_b|v_a\rangle=\delta_{ab},\; \langle v_b|v_0\rangle=0$.
Its eigenvalues are
\begin{equation}\label{}
  \epsilon:=\frac{N^2-1}{N^2},\; \epsilon_1:=-\frac{1}{N^2}
\end{equation}
with
$  \mathcal{M}|v_0\rangle=\epsilon|v_0\rangle,
  \mathcal{M} |v_a\rangle= \epsilon_1|v_a\rangle, \; a=1,\dots,N^2-1$.
The normalization of ground states are
\begin{equation}\label{}
  \langle \mathbf{L}|\mathbf{L}\rangle=\text{tr}\mathcal{M}^L/\text{tr}\mathcal{M}^L=1,
  \langle \mathbf{R}|\mathbf{R}\rangle=\text{tr}(\mathcal{M}^t)^L/\text{tr}(\mathcal{M}^t)^L=1,
\end{equation}
for $\mathcal{M}^t:=\frac{2}{N}\sum_m t^{m*}\otimes t^{m}=(\mathcal{M})^t$
as the transpose of $\mathcal{M}$.
The overlap
$\langle \mathbf{L}|\mathbf{R}\rangle = \text{tr}\mathcal{N}^L/\text{tr}\mathcal{M}^L$
for
\begin{equation}\label{}
  \mathcal{N}:=\frac{2}{N}\sum_m t^m\otimes t^{m}.
\end{equation}
Define
$  |n_a\rangle :=\sqrt{2}\sum_{ij}t^{a*}_{ij}|ij\rangle$,
then
$  \langle n_b|n_a\rangle=\delta_{ab},\; \langle \omega|n_a\rangle=0$.
In the set of Gell-Mann matrices $\{t^m\}$,
there are $N(N-1)/2$ symmetric (sym)
and $N(N-1)/2$ anti-symmetric (anti) matrices,
and $N-1$ diagonal matrices in the Cartan subalgebra.
We find

\begin{align}\label{}
    \langle n_b|v_a\rangle=\left\{
                \begin{array}{ll}
                  \delta_{ab}, & (\text{sym,\;Cartan}),\\
                  -\delta_{ab}, &  (\text{anti}).
                \end{array}
              \right.\;
  \langle n_a|\mathcal{N}|n_a\rangle=\left\{
                \begin{array}{ll}
                \frac{N-1}{N^2}, & (\text{sym,\;Cartan}),\\
                 -\frac{N+1}{N^2}, & (\text{anti}).
                 \end{array}
              \right.
  \end{align}
Together with $\langle \omega|\mathcal{N}|\omega\rangle=\frac{N-1}{N^2}$,
this means $\mathcal{N}$ has
$N(N-1)/2$ degenerate eigenvalues $-\frac{N+1}{N^2}$,
and $N(N+1)/2$ degenerate eigenvalues $\frac{N-1}{N^2}$,
and together they satisfy $\text{tr}\mathcal{N}=0.$
Therefore,
\begin{equation}\label{}
  \langle \mathbf{L}|\mathbf{R}\rangle = \text{tr}\mathcal{N}^L/\text{tr}\mathcal{M}^L
  =\left(\frac{1}{N-1}\right)^L\rightarrow 0.
\end{equation}

To compute energy, we define
\begin{align}
  \mathcal{M}(T^\alpha)&:=-i\frac{2}{N} \sum_{mn} f_{\alpha mn} t^n\otimes t^{m*},\\
    \mathcal{M}(T^\alpha T^\beta)& :=-\frac{2}{N} \sum_{mnk} f_{\alpha mn} f_{\beta nk} t^k\otimes t^{m*},
\end{align}
for the generators $T^\alpha =-i\sum_{mn} f_{\alpha mn}|m\rangle\langle n|$
with structure constants $\{f_{\alpha mn}\}$ of $SU(N)$.
We find the following properties
\begin{align}\label{}
(\mathcal{M}(T^\alpha))^\dagger &=-\mathcal{M}(T^\alpha),\;
\mathcal{M}^t(T^\alpha) = (\mathcal{M}(T^\alpha))^t,\;
\mathcal{M}^t((T^\alpha)^2) = (\mathcal{M}((T^\alpha)^2))^t,\\
(\mathcal{M}(T^\alpha T^\beta))^\dagger&=  \mathcal{M}(T^\alpha T^\beta) ,
  \mathcal{M}^t(T^\alpha T^\beta)= (\mathcal{M}(T^\alpha T^\beta))^t,
\end{align}
and with easy algebra we find
\begin{align}\label{eq:energygs}
  \langle \mathbf{L}|h_n|\mathbf{L}\rangle
  &=\sum_\alpha \text{tr} \mathcal{M}^{L-2} \mathcal{M}(T^\alpha)^2 / \text{tr} \mathcal{M}^L
  =-\frac{N^3}{2(N^2-1)}, \; \forall n,\\
  \langle \mathbf{L}|h_n^2|\mathbf{L}\rangle
  &= \sum_{\alpha\beta} \text{tr} \mathcal{M}^{L-2} \mathcal{M}(T^\alpha T^\beta)^2/\text{tr} \mathcal{M}^L
  =\frac{N^2(N^2+2)}{4(N^2-1)}, \; \forall n,
\end{align}
and also $\langle \mathbf{R}|h_n|\mathbf{R}\rangle=\langle \mathbf{L}|h_n|\mathbf{L}\rangle$,
$ \langle \mathbf{R}|h_n^2|\mathbf{R}\rangle = \langle \mathbf{L}|h_n^2|\mathbf{L}\rangle $.
As a result,
\begin{equation}\label{}
  \langle \mathbf{L}|H_n|\mathbf{L}\rangle= \langle \mathbf{R}|H_n|\mathbf{R}\rangle=0, \; \forall n,
\end{equation}
as expected.

The system is gapped and has an exponentially-decaying
correlation function~\cite{GR07,KHK08,RSS+10,OT11,MUM+14}.
For a general ground state $|\mathbf{G}\rangle\in \{|\mathbf{L}\rangle, |\mathbf{R}\rangle\}$,
we confirmed that for single-site observable $T^\alpha_n$ it holds
\begin{align}\label{}
  \langle\mathbf{G}| T^\alpha |\mathbf{G}\rangle =0,
  \langle\mathbf{G}| (T^\alpha)^2 |\mathbf{G}\rangle
  =\frac{N}{N^2-1}.
\end{align}
This means the net ``polarization'' $\sum_n T^\alpha_n$ of the ground states is zero,
yet there are fluctuations $\sum_n (T^\alpha_n)^2$.

For correlation function we find
\begin{align}\label{}
  \langle\mathbf{G}|T_m^\alpha T_n^\beta|\mathbf{G} \rangle
  = \frac{N^3}{2(N^2-1)} \left( \frac{-1}{N^2-1} \right)^r\delta_{\alpha\beta}
\end{align}
for $r:=|n-m|$ as the spatial distance.
For $r=1$, this is consistent with the ground state value of $h_n$
(see Eq.~(\ref{eq:energygs})).

Now we consider excitations.
First, a singlet bond can be excited to be in the adjoint irrep,
termed an \emph{adjointor}.
For $SU(2)$ case, this excitation is the so-called ``triplon''~\cite{Kna88,FS93,TS95}.
An adjointor from $|\mathbf{L}\rangle$ is modelled by changing
a bond $|\omega\rangle$ to $|v_\alpha\rangle$ for any $\alpha$
(see Eq.~(\ref{eq:basiss})).
With this change between site $n$ and $n+1$,
we get an adjointor state
\begin{equation}\label{}
  |\mathbf{A}(\alpha, n)\rangle:=
  \vartheta\sqrt{2N}\sum_{i_1\cdots i_L}
  \text{tr} (A^{i_1}\cdots A^{i_n} t^\alpha A^{i_{n+1}}\cdots A^{i_L} ) |i_1\cdots i_L\rangle.
\end{equation}
Recall that $\vartheta=\left(\frac{N^2}{N^2-1}\right)^{L/2}$.
We find
\begin{align}\label{eq:virtualcorr}
  \langle\mathbf{A}(\beta,m)|\mathbf{A}(\alpha, n)\rangle&=
  \left(\frac{-1}{N^2-1}\right)^{|n-m|} \delta_{\alpha\beta}.
\end{align}
For its energy, with
\begin{align}\label{eq:adjhE}
  \langle \psi(\beta, n)|h_n |\psi(\alpha, n)\rangle =
\frac{N^3}{2(N^2-1)^2} \delta_{\alpha\beta},\;
  \langle \psi(\beta, n)|h_n^2 |\psi(\alpha, n)\rangle =
\frac{N^2(3N^2-2)}{4(N^2-1)^2} \delta_{\alpha\beta},
\end{align}
we find
\begin{equation}\label{}
  \langle\mathbf{A}(\beta,m)|H_n|\mathbf{A}(\alpha, n)\rangle=
  \frac{N^3(N^2+1)}{3(N^2-1)^2} \delta_{\alpha\beta}\delta_{mn},
\end{equation}
which is $\frac{N}{3}$ in the large-$N$ limit.

The energy of dimer states can also be straightforwardly obtained.
For simplicity we assume there are even number of sites so there will be no free bonds.
For a chain with $L$ sites, there are $L/2$ pairs with two bonds for each pair.
A dimer state can be written as
\begin{equation}\label{}
  |\mathbf{D}\rangle=\vartheta_\mathbf{D} \left(\sum_{i_l,i_r} \text{tr}(A^{i_l}A^{i_r})|i_l i_r\rangle \right)^{\otimes L/2}
\end{equation}
with normalization constant $\vartheta_\mathbf{D}:=(\frac{N^2}{N^2-1})^{L/4}$.
To compute the energy of a local term $h$ (the site label is ignored), there are two cases:
(I) $h$ acts on a pair with two bonds in between;
(II) $h$ acts on a pair with no bond in between.

For case (I), the energy is
\begin{equation}\label{}
  \langle\mathbf{D}|h|\mathbf{D}\rangle
  =\sum_\alpha \text{tr} \mathcal{M}(T^\alpha)^2 /\text{tr} \mathcal{M}^2=-N,
\end{equation}
which is lower than the value on ground states.
For $h^2$, the energy is
\begin{equation}\label{}
  \langle \mathbf{D}|h^2|\mathbf{D}\rangle
  =\sum_{\alpha\beta} \text{tr} \mathcal{M}(T^\alpha T^\beta)^2
   /\text{tr} \mathcal{M}^2
  =N^2,
\end{equation}
which is higher than the value on ground states.
However, the energy of $H_n$ is $-N+\frac{2}{3N}N^2+\frac{N}{3}=0$,
the same with ground state energy.

For case (II), we see that there is no entanglement between the pair,
and the reduced state of them is a totally mixed state.
We find the energy of $h$ is zero, of $h^2$ is $\frac{N^2}{N^2-1}$,
so the energy of $H_n$ is $\frac{N(N^2+1)}{3(N^2-1)}$,
higher than that on ground states.

As the result, for a ring of length $L$, the total energy of a dimer state is
\begin{equation}\label{}
  \langle \mathbf{D}|H_N|\mathbf{D}\rangle=L\frac{N(N^2+1)}{6(N^2-1)}.
\end{equation}
In the large-$N$ limit, this value for each bond is $N/6$.

\subsection{Parity symmetry}
\label{sec:parity}

Here we prove the parity symmetry of the model
which is spontaneously broken.
For the ground states represented as matrix-product states,
instead of Gell-Mann matrices $\{t^m\}$
we can also use the set of $E_{ij}:=|i\rangle\langle j|$ for $i\neq j$
and the set of diagonal Gell-Mann matrices,
denoted by $\{E_k\}$ in this section.
There are in total $N^2-N$ matrices $E_{ij}$ and $N-1$ matrices $E_k$.
For the ground state $|\mathbf{L}\rangle$,
the set of matrices on each site are
\begin{equation}\label{}
  E_{12},E_{21},E_{13},E_{31},\cdots, E_{N-1,N}, E_{N,N-1},E_1,\cdots, E_{N-1},
\end{equation}
and for $|\mathbf{R}\rangle$ they are
\begin{equation}\label{}
  E_{21},E_{12},E_{31},E_{13},\cdots, E_{N,N-1}, E_{N-1,N}, E_1,\cdots, E_{N-1}.
\end{equation}
With the matrices $E_{ij}$ and $E_k$,
we denote the on-site physical basis as $\{\{|E_{ij}\rangle\},\{|E_k\rangle\}\}$,
and then the on-site permutation $\Pi$ can be formally expressed as
\begin{equation}\label{}
  \Pi= \sum_k |E_k\rangle\langle E_k|
     + \sum_{ij} |E_{ij}\rangle\langle E_{ji}|+|E_{ji}\rangle\langle E_{ij}|.
\end{equation}
Now the action of the logical $\bar{X}$ on the two states are apparent,
which is to switch between $E_{ij}$ and $E_{ji}$.

Furthermore, the Hamiltonian $H_N$ has the parity symmetry due to the following property
\begin{equation}\label{}
  (\Pi\otimes\Pi) h_n (\Pi\otimes\Pi) =h_n.
\end{equation}
This is proved as follows.
In the adjoint irrep,
we denote the corresponding matrices for $E_{ij}$ as $\mathscr{E}_{ij}$,
and for $E_k$ as $\mathscr{E}_{k}$.
The term $h_n$ can be formally expressed as
\begin{equation}\label{}
  h_n =\frac{1}{2}\sum_{ij} \mathscr{E}_{ij} \otimes \mathscr{E}_{ji}
    + \sum_k \mathscr{E}_{k} \otimes \mathscr{E}_{k}.
\end{equation}
In the basis $\{\{|E_{ij}\rangle\},\{|E_k\rangle\}\}$,
an entry of $\mathscr{E}_{k}$ is denoted like $\mathscr{E}_{k}(E_{ij}, E_{ij})$,
and we find that $\mathscr{E}_{k}$ are all diagonal and
\begin{align}\label{}
 \mathscr{E}_{k}(E_{ij}, E_{ij}) =\text{tr} E_{ji}(E_k E_{ij}- E_{ij} E_k)=-\mathscr{E}_{k}(E_{ji}, E_{ji}),\;
 \mathscr{E}_{k}(E_i, E_i) =0.
\end{align}
Then
\begin{equation}\label{}
  \Pi \mathscr{E}_k \Pi= -\mathscr{E}_k.
\end{equation}
Also with
\begin{align}\label{}
  \mathscr{E}_{ij}(E_\alpha,E_\beta) =& \text{tr} E_\alpha(E_{ij}E_\beta-E_\beta E_{ij})=0, \\
  \mathscr{E}_{ij}(E_{xy},E_\beta) =& \text{tr} E_{yx} (E_{ij}E_\beta-E_\beta E_{ij})=0,
\end{align}
we find
\begin{equation}\label{}
    \mathscr{E}_{ij} =\sum_\ell |E_{i\ell}\rangle\langle E_{j\ell}|-|E_{\ell j}\rangle\langle E_{\ell i}|
    =(\mathscr{E}_{ji})^t.
\end{equation}
Then
\begin{equation}\label{}
  \Pi \mathscr{E}_{ij} \Pi= -\mathscr{E}_{ji}.
\end{equation}
As $h_n$ is a two-body interaction,
now it is clear to see that
the Hamiltonian $H_N$ is invariant under the global permutation.

\renewcommand{\theequation}{SB\arabic{equation}}
\renewcommand{\thefigure}{SB\arabic{figure}}
\renewcommand{\thetable}{SB\arabic{table}}

\section{Twist operation}
\label{sec:ztwist}

In this section we present a detailed study of the twist operation
that serves as a topological logical $Z$-rotation gate.
These include the effects of twist on ground states, excitations, the energy cost
of twist, and a representation of twist by fermionic operators.

A general twist operator is defined as
\begin{equation}\label{}
  U_\textsc{tw}(\{\theta_n\},f):=\otimes_{n=1}^L e^{i\theta_n \mathcal{O}_n}, \;
  \sum_n (\theta_n-\theta_{n-1})=2\pi f,\;
  \theta_n-\theta_{n-1} \in O(1/L),\; f\in[0,1],
\end{equation}
for a diagonal operator $\mathcal{O}$ in Cartan subalgebra
and $e^{i2\pi \mathcal{O}}=\mathds{1}$.
The ``uniform'', or undisturbed twist is
\begin{equation}\label{}
  U_\textsc{tw}(f):=\otimes_{n=1}^L e^{i\ell n f \mathcal{O}_n}, \;
\ell:=2\pi/L.
\end{equation}
Note that twist operators are often employed in the context of
Lieb-Schultz-Mattis theorem~\cite{LSM61,AL86}.
We will term the full twist at $f=1$ simply as the twist $U_\textsc{tw}$,
and $U_\textsc{tw}(\frac{1}{2})$ as the half-twist,
and other cases as ``fractional'' twist.
It is more convenient to study the twist effect in the virtual space picture.
Let $U_n:=e^{i\ell n \mathcal{O}_n}$,
then the $SU(N)$ symmetry implies there exist $V_n$ such that
\begin{equation}\label{}
  \sum_j \langle i|U_n|j\rangle A^j = V_n A^i V_n^\dagger, \; U_n=(U_1)^n,\; V_n=(V_1)^n.
\end{equation}
We will denote $V_1$ simply as $V$,
and choose $V=\text{diag}(e^{i\ell},1,\dots,1)$ below,
while the twist effects also hold for other flavors of $V$.
Recall that flavors refer to the positions on the diagonal of $V$.

\subsection{Twist on ground states}
\label{sec:ztwist0}

Here we use the transfer matrix method to
find the twist effect.
Using the matrices $\{E_{ij}\}$ and diagonal Gell-Mann matrices $\{E_k\}$,
as in section~\ref{sec:parity}, we find
\begin{align}\label{}
\begin{split}
  \mathcal{M} =&\frac{1}{N} (E_{12}\otimes E_{12} + \dots+ E_{1N}\otimes E_{1N})+ \\
   &\frac{1}{N} (E_{21}\otimes E_{21} +  \dots+E_{2N}\otimes E_{2N})+\dots \\
   &\frac{1}{N} (E_{N1}\otimes E_{N1} +  \dots+E_{N,N-1}\otimes E_{N,N-1})+ \\
   &\frac{2}{N}(E_1\otimes E_1 + \cdots +E_{N-1}\otimes E_{N-1}),
\end{split}
\end{align}
\begin{align}\label{}
\begin{split}
  \mathcal{N} =&\frac{1}{N} (E_{12}\otimes E_{21} + \dots+ E_{1N}\otimes E_{N1})+ \\
   &\frac{1}{N} (E_{21}\otimes E_{12} +  \dots+E_{2N}\otimes E_{N2})+\dots \\
   &\frac{1}{N} (E_{N1}\otimes E_{1N} +  \dots+E_{N,N-1}\otimes E_{N-1,N})+ \\
   &\frac{2}{N}(E_1\otimes E_1 + \cdots +E_{N-1}\otimes E_{N-1}).
\end{split}
\end{align}
Define $\mathcal{G}:=\mathcal{V}\mathcal{M}$ with $\mathcal{V}=V\otimes\mathds{1}$ for
$V=\text{diag}(e^{i\ell},1,\dots,1)$.
Notice that the twist works equally well for other flavors of $V$
when the phase $e^{i\ell}$ is at other positions on the diagonal of $V$,
here we assume it is on the first place.
The effect of $V$ is to change $E_{1b}$ to $e^{i\ell}E_{1b}$, and
the effects on the $E_k$ is clear from its action on
the maximally entangled state $|\omega\rangle$,
which is expressed as
$|\omega\rangle=\frac{1}{\sqrt{N}}(|11\rangle+\cdots |NN\rangle)$
in this section.
The $|11\rangle$ component obtains a phase $e^{i\ell}$,
while the other parts do not change.
Formally,
\begin{align}\label{}
\begin{split}
  \mathcal{G} =&\frac{e^{i\ell}}{N} (E_{12}\otimes E_{12} + \dots+ E_{1N}\otimes E_{1N})+ \\
   &\frac{1}{N} (E_{21}\otimes E_{21} + \dots+ E_{2N}\otimes E_{2N})+\dots \\
   &\frac{1}{N} (E_{N1}\otimes E_{N1} + \dots+ E_{N,N-1}\otimes E_{N,N-1})+ \\
   &\frac{2}{N}(\tilde{E}_1\otimes E_1 + \cdots +\tilde{E}_{N-1}\otimes E_{N-1}),
\end{split}
\end{align}
for $\tilde{E}_i$ denoting the properly modified $E_i$.
With the properties
\begin{align}\label{}
  E_{ij} \otimes E_{ij} |\omega\rangle =\frac{1}{\sqrt{N}}|ii\rangle, \;
  \sum_k E_k\otimes E_k  |\omega\rangle =\frac{N-1}{2N}|\omega\rangle,
\end{align}
and denote $|22\rangle+\cdots |NN\rangle:= |\bar{11}\rangle$,
we find
\begin{align}\label{}
\begin{split}
  \mathcal{G}|\omega\rangle&= e^{i\ell} \frac{N-1}{N\sqrt{N}}|11\rangle
  + \frac{N-1}{N\sqrt{N}}|\bar{11}\rangle + e^{i\ell} \frac{N-1}{N^2\sqrt{N}}|11\rangle
  + \frac{N-1}{N^2\sqrt{N}}|\bar{11}\rangle,\\
  \mathcal{G}|11\rangle&= \frac{1}{N}|\bar{11}\rangle+ \frac{N-1}{N^2}|11\rangle.
\end{split}
\end{align}
Therefore,
the eigenvector of $\mathcal{G}$ with the largest eigenvalue magnitude
is a linear combination of $|11\rangle$ and $|\bar{11}\rangle$,
so
\begin{equation}\label{}
  \mathcal{G} (\alpha |11\rangle +\beta|\bar{11}\rangle) =\lambda (\alpha |11\rangle +\beta|\bar{11}\rangle).
\end{equation}
We find
\begin{equation}\label{}
  2(\lambda-\frac{N^2-N-1}{2N})\beta =\alpha,
\end{equation}
and $\lambda$ is the solution of the following equation
\begin{equation}\label{}
  4N^2 \lambda^2-2N(N^2-2)\lambda+(N^2-1)(N-1)-N(N^2-1)e^{i\ell}=0.
\end{equation}
With the approximation $e^{i\ell}\approx 1+ i\ell$ we find
\begin{equation}\label{}
  \lambda= \frac{N^2-1}{N^2}e^{i\ell/N},\;
  \beta=\frac{1}{\sqrt{N}},\;
  \alpha=e^{i\ell\frac{N^2-1}{N^2}}\beta.
\end{equation}
In the large-$L$ limit and for $\ell=2\pi/L$, we find
\begin{equation}\label{eq:twsit}
  \langle \mathbf{L}|U_\textsc{tw}|\mathbf{L}\rangle=\text{tr}\mathcal{G}^L/\text{tr}\mathcal{M}^L=e^{i2\pi/N},\; \langle \mathbf{R}|U_\textsc{tw}|\mathbf{R}\rangle=e^{-i2\pi/N},
\end{equation}
which is one of our central results.

In addition, the value of $\text{tr}\mathcal{G}^L$ can also be computed as
$(\langle\omega|\mathcal{G}|\omega\rangle)^L$.
With
\begin{equation}\label{}
  \langle\omega|\mathcal{G}|\omega\rangle=\frac{N^2-1}{N^2}\frac{N-1+e^{i\ell}}{N},
\end{equation}
we find the same results~(\ref{eq:twsit}).

Also to compute $\langle \mathbf{L}|U_\textsc{tw}|\mathbf{R}\rangle$, we find
\begin{equation}\label{}
  \langle \mathbf{L}|U_\textsc{tw}|\mathbf{R}\rangle = \text{tr}\mathcal{P}^L/\text{tr}\mathcal{M}^L
\end{equation}
for $\mathcal{P}:=\mathcal{V}\mathcal{N}$ and formally,
\begin{align}\label{}
\begin{split}
  \mathcal{P} =&\frac{e^{i\ell}}{N} (E_{12}\otimes E_{21} + \dots+ E_{1N}\otimes E_{N1})+ \\
   &\frac{1}{N} (E_{21}\otimes E_{12} + \dots+ E_{2N}\otimes E_{N2})+\dots \\
   &\frac{1}{N} (E_{N1}\otimes E_{1N} + \dots+ E_{N,N-1}\otimes E_{N-1，N})+ \\
   &\frac{2}{N}(\tilde{E}_1\otimes E_1 + \cdots +\tilde{E}_{N-1}\otimes E_{N-1}).
\end{split}
\end{align}
For the trace $\text{tr}\mathcal{P}^L$, we only need to consider the eigenvectors
$|n_a\rangle$ of $\mathcal{N}$
for the antisymmetric case.
To make this clearer, we denote an antisymmetric Gell-Mann matrix as
$t^{ab}=i(E_{ba}-E_{ab})/2$,
then $|n_{ab}\rangle=\frac{\sqrt{2}}{2}(|ba\rangle-|ab\rangle)$.
From $\mathcal{P}|n_{ab}\rangle=-\frac{N+1}{2N}|n_{ab}\rangle$, we find
\begin{equation}\label{}
  \mathcal{P}|ab\rangle=\frac{1}{2}|ba\rangle-\frac{e^{i\ell}}{2N}|ab\rangle,
  \mathcal{P}|ba\rangle=\frac{e^{i\ell}}{2}|ab\rangle-\frac{1}{2N}|ba\rangle,
\end{equation}
so the superposition of $|ab\rangle$ and $|ba\rangle$ is an eigenvector
\begin{equation}\label{}
  \mathcal{P}(\alpha |ab\rangle+\beta |ba\rangle) =\lambda (\alpha |ab\rangle+\beta |ba\rangle),
\end{equation}
and we find
\begin{equation}\label{}
  \lambda=-\frac{N+1}{N^2}e^{i\ell/2}, \beta=1/\sqrt{2}, \alpha=-e^{i\frac{N+1}{N^2}\ell}\beta.
\end{equation}
We arrive at
\begin{equation}\label{}
  \langle \mathbf{L}|U_\textsc{tw}|\mathbf{R}\rangle = \left(\frac{1}{N-1}\right)^L e^{i\ell L/2}\rightarrow 0.
\end{equation}

Next we analyze the effect of a fractional twist
if we replace $\ell$ by $\ell'=f \ell$ for a fractional real number $f\in [0,1]$,
which can be viewed as a time parameter.
The overlap is
\begin{equation}\label{}
  \langle \mathbf{L}|U_\textsc{tw}(f)|\mathbf{L}\rangle
  =\epsilon_*^L \langle \epsilon| \tilde{\Lambda}|\epsilon\rangle/\text{tr}\mathcal{M}^L
\end{equation}
for $\tilde{\Lambda}=\Lambda\otimes \mathds{1}$,
$\Lambda=\text{diag}(e^{-i2\pi f},1,\dots,1)$,
$\epsilon_*$ as the largest eigenvalue of $\mathcal{G}$
and $|\epsilon\rangle$ the corresponding eigenvector.
Note here $\mathcal{G}$ is defined for $\ell'$ instead of $\ell$.
We find
$  \langle \epsilon| \tilde{\Lambda} |\epsilon\rangle=(N-1+e^{-i2\pi f})/N$,
so
\begin{align}\label{}
  \langle \mathbf{L}|U_\textsc{tw}(f)|\mathbf{L}\rangle=\frac{1}{N} e^{i2\pi f/N} (N-1+e^{-i2\pi f}),\;
  \langle \mathbf{R}|U_\textsc{tw}(f)|\mathbf{R}\rangle=\frac{1}{N} e^{-i2\pi f/N} (N-1+e^{i2\pi f}),
\end{align}
and the off-diagonal elements vanish.
If $f$ is very small, we can see that
$\langle \mathbf{L}|U_\textsc{tw}|\mathbf{L}\rangle
=\langle \mathbf{R}|U_\textsc{tw}|\mathbf{R}\rangle=1$,
which basically means that the twist has trivial effect on the states.
If $f$ is slightly perturbed from 1 by $\delta f$, i.e.,
$f+|\delta f|=1$, $|\delta f|\ll 1$, the twist angle $e^{\pm i2\pi/N}$
remains the same with corrections of the order $(\delta f)^2$.
This means the phase gate is stable against the perturbations by $\delta f$.
Also for non-small $f$, the ground state will be excited a little.
This means we cannot do phase rotations for arbitrary angles beyond $e^{i2\pi/N}$,
which is a topological feature of the state.
However, for the half-twist at $f=\frac{1}{2}$,
we have
\begin{align}\label{}
  \langle \mathbf{L}|U_\textsc{tw}(\frac{1}{2})|\mathbf{L}\rangle=\frac{N-2}{N} e^{i\pi/N},\;
  \langle \mathbf{R}|U_\textsc{tw}(\frac{1}{2})|\mathbf{R}\rangle=\frac{N-2}{N} e^{-i\pi/N}.
\end{align}
The factor $\frac{N-2}{N}$ means there will be excitations.
However, this factor is the same for the two ground states,
which means that such excitation does not induce any logical error.
By cooling we can maintain in the code space,
hence the half-twist $U_\textsc{tw}(\frac{1}{2})$
can be properly denoted as $\sqrt{U_\textsc{tw}}$.

\begin{example}
Now we show an example for the $SU(3)$ case.
With a general phase gate
\begin{equation}\label{}
  V=\emph{diag}(1,b,c), \; b,c\in U(1)
\end{equation}
acting on the virtual space,
its on-site adjoint irrep $U$ takes the form
\begin{equation}\label{}
  U=\emph{diag}(1/b,b,c,1/c,b/c,c/b,1,1),\; b,c\in U(1).
\end{equation}
There are three flavors $u,d,s$ (for up, down, strange) in the virtual space.
For the three operators
\begin{equation}\label{}
  V_u=\emph{diag}(e^{i\ell},1,1),\;
  V_d=\emph{diag}(1,e^{i\ell},1),\;
  V_s=\emph{diag}(1,1,e^{i\ell}),
\end{equation}
we find
\begin{align}\label{}
  U_u&=\emph{diag}(e^{i\ell},e^{-i\ell},e^{-i\ell},e^{i\ell},1,1,1,1),\\
  U_d&=\emph{diag}(e^{-i\ell},e^{i\ell},1,1,e^{i\ell},e^{-i\ell},1,1),\\
  U_s&=\emph{diag}(1,1,e^{i\ell},e^{-i\ell},e^{-i\ell},e^{i\ell},1,1),
\end{align}
for $\ell:= 2\pi/L$.
Also $U_f=e^{i\ell \mathcal{O}_f}$ ($f=u,d,s$) for
\begin{align}\label{}
  \mathcal{O}_u  =T^3+ T^8/\sqrt{3},\;
  \mathcal{O}_d  =-T^3+ T^8/\sqrt{3},\;
    \mathcal{O}_s  =-2T^8/\sqrt{3},
  \end{align}
with Gell-Mann matrices $T^3$ and $T^8$ in the adjoint irrep.
Denote the twist as $U_\textsc{tw}$,
on the two ground states we find
\begin{align}\label{}
  \langle \mathbf{L}| U_\textsc{tw}|\mathbf{L }\rangle = e^{i2\pi/3},\;
  \langle \mathbf{R}| U_\textsc{tw}|\mathbf{R}\rangle = e^{-i2\pi/3},\;
  \langle \mathbf{L}| U_\textsc{tw}|\mathbf{R}\rangle=0.
  \end{align}
So the logical gate is
\begin{equation}\label{}
  \bar{Z}(2\pi/3)=\begin{pmatrix} 1 & 0 \\ 0 & e^{i2\pi/3} \end{pmatrix}.
\end{equation}
which is of order three such that $(\bar{Z}(e^{i2\pi/3}))^3=\mathds{1}$.
In addition, from the half-twist we obtain another logical gate as
\begin{equation}
  \bar{Z}(-2\pi/3)=\begin{pmatrix} 1 & 0 \\  0 & e^{-i2\pi/3}  \end{pmatrix},
\end{equation}
which is also of order three such that $(\bar{Z}(e^{-i2\pi/3}))^3=\mathds{1}$.\\
Furthermore, to manifest the difference between even and odd $N$ cases,
for $SU(4)$ the full twist leads to the logical gate
\begin{equation}\label{}
  \bar{Z}(\pi)=\begin{pmatrix} 1 & 0 \\ 0 & e^{i\pi} \end{pmatrix}=\sigma_z,
\end{equation}
which is the Pauli $\sigma_z$,
while the half-twist leads to the logical gate
\begin{equation}
  \bar{Z}(-\pi/2)= \begin{pmatrix} 1 & 0 \\ 0 & e^{-i\pi/2} \end{pmatrix},
\end{equation}
which is equivalent to the phase gate
$S=\left(\begin{smallmatrix} 1 & 0 \\ 0 & i \end{smallmatrix}\right)$ and of order four.
\end{example}

\begin{remark}\label{remark:homotopy}
  The twist operation can also be defined on a segment of the system.
  When the twist acts on a long-enough continuous segment with length $\tilde{L}$,
  the phase factor we obtain on $|\mathbf{L}\rangle$ is
  $(\frac{e^{i\tilde{\ell}}+N-1}{N})^{\tilde{L}}$ for $\tilde{\ell}=\frac{2\pi}{\tilde{L}}$,
  which is still $e^{i2\pi/N}$.
  Furthermore, the value of $\tilde{\ell}$ can also be different on each bond
  as long as $\sum_b \tilde{\ell}_b =2\pi$
  and the deviation from $\frac{2\pi}{\tilde{L}}$ is small.
  This means that the twist operation is homotopic:
  the twist phase can be accumulated in different ways
  as long as the sum of those phase factors is $2\pi$.
\end{remark}

\subsection{Twist via Fermion representation}
\label{sec:qft}

Here we present a fermion representation of the system
and analyze the twist effect.
The effect of twist can be well characterized using the fermion operator representation.
With Young tableau~\cite{Ram10} in mind, we label
the fundamental irrep of $SU(N)$ as $\blacksquare$, treated as a particle,
and its conjugate can be treated as a hole, denoted as $\Box$,
which is equivalent to $N-1$ black boxes in one column.
In this section, Einstein's summation rule is assumed.

For $\blacksquare$, its $SU(N)$ generators are
$S^i_j= \psi^{\dagger i}\psi_j -\delta^i_j/N$ for
$S^i_i=0$ and $\psi^{\dagger i}\psi_i=1$, $i, j=1,2,\dots, N$.
For $\Box$, its $SU(N)$ generators are
$S^i_j=-\bar{\psi}^\dagger_j \bar{\psi}^i + \delta^i_j/N$
for $S^i_i=0$ and $\bar{\psi}^{\dagger}_i\bar{\psi}^i=1$
with the hole operator $\bar{\psi}^i:=\psi^{\dagger i}$~\cite{Aff85}.
For the adjoint irrep $\Box\blacksquare$, we define its generators as
\begin{equation}\label{}
  S^i_j:=\psi^{\dagger i}\psi_{j}- \bar{\psi}^{\dagger}_j \bar{\psi}^i.
\end{equation}
The ground states are each a product of singlets
\begin{equation}\label{}
  |\mathbf{L}\rangle=\prod_{n=1}^L \bar{\psi}_i^\dagger(n) \psi^{\dagger i}(n+1) |F,\Omega\rangle^{\otimes L},\;
  |\mathbf{R}\rangle=\prod_{n=1}^L \psi^{\dagger i}(n) \bar{\psi}_i^\dagger(n+1) |F,\Omega\rangle^{\otimes L},
\end{equation}
for $|\Omega\rangle$ as the vacuum of particle
and $|F\rangle$ as the vacuum of hole.
For the PBC case, there are $L$ bonds and each bond has $N$ flavors $f_i$,
then each ground state can be formally written as
\begin{equation}\label{}
\frac{1}{\sqrt{N^L}}\sum_{i_1,\dots,i_L}|f_{i_1}\cdots f_{i_L}\rangle.
\end{equation}

The twist only changes a flavor, say $\alpha$,
from $\bar{\psi}_\alpha^\dagger(n) \psi^{\dagger \alpha}(n+1)$
to $e^{i\ell}\bar{\psi}_\alpha^\dagger(n) \psi^{\dagger \alpha}(n+1)$,
and
from $\psi^{\dagger \alpha}(n) \bar{\psi}_\alpha^\dagger(n+1)$
to $e^{-i\ell}\psi^{\dagger \alpha}(n)\bar{\psi}_\alpha^\dagger(n+1)$.
Then for $|\mathbf{L}\rangle$ after the twist we can see that it becomes
\begin{equation}\label{}
|\mathbf{L}'\rangle=\frac{1}{\sqrt{N^L}}\sum_{i_1,\dots,i_L} e^{i\ell \sum_{n=1}^L \delta_{f_n,\alpha} } |f_{i_1}\cdots f_{i_L}\rangle,
\end{equation}
and the overlaps are
\begin{equation}\label{}
  \langle \mathbf{L}|\mathbf{L}'\rangle= (\frac{N-1+e^{i\ell}}{N})^L,\;
  \langle \mathbf{R}|\mathbf{R}'\rangle= (\frac{N-1+e^{-i\ell}}{N})^L.
\end{equation}
For small $\ell=2\pi/L$, this leads to
\begin{equation}\label{}
  \langle \mathbf{L}|\mathbf{L}'\rangle=e^{i2\pi/N},\; \langle \mathbf{R}|\mathbf{R}'\rangle=e^{-i2\pi/N},
\end{equation}
the same with the results obtained using the transfer matrix method.
Note that if the parameter is $\ell f$ for a fractional number $f$,
then there will be an additional term in the new ground state,
as has been studied in the section~\ref{sec:ztwist0}.

\subsection{Energy cost of twist}
\label{sec:z-cost}

We have seen that the fractional twist will excite the system.
Here we compute the energy cost of the twist as a function of $f$.
This also means the twist is protected by the gap of the system.

For $h_n$ on sites $n$ and $n+1$,
there are two cases:
(I) $\Lambda\otimes \Lambda^*$ is not in between $n$ and $n+1$;
(II) $\Lambda\otimes \Lambda^*$ is in between $n$ and $n+1$.
For case (I),
the energy is
\begin{align}\label{}
  E_\textsc{i}:=-\sum_\alpha (\frac{N^2}{N^2-1})^2 \text{tr}(t^\alpha V^\dagger t^\alpha V)/N
\end{align}
for $\mathcal{R}=(V\otimes V^*)\mathcal{M}$.
For case (II), the energy is
\begin{align}\label{}
E_\textsc{ii}:=-\sum_\alpha (\frac{N^2}{N^2-1})^2 \text{tr}(t^\alpha V^\dagger \Lambda^\dagger t^\alpha V\Lambda )/N.
\end{align}
We expand $V$ and $W\equiv V\Lambda$ for the twist in terms of Pauli matrices and identity, and we find $V=v_0\mathds{1}+\sum_{i=1}^{N-1} v_i Z^i $
and $W=w_0\mathds{1}+\sum_{i=1}^{N-1} w_i Z^i $ for
\begin{align}\label{}
  v_0&=(1-1/N)+e^{i\ell f}/N, \;\; v_i=(e^{i\ell f}-1)/N, \\
  w_0&=(1-1/N)+e^{i\ell f(1-L)}/N, w_i=(e^{i\ell f(1-L)}-1)/N.
\end{align}
So we find
\begin{align}\label{}
E_\textsc{i}=-\frac{N^3}{2(N^2-1)},\;
E_\textsc{ii}=\frac{N^3}{2(N^2-1)} (-|w_0|^2 +\frac{1}{N^2-1} \sum_\ell |w_\ell|^2).
\end{align}
From $\frac{\partial E_\textsc{ii}}{\partial f}=0$,
we find that $E_\textsc{ii}$ takes the maximum at $f=1/2$.
At $f=1/2$, $w_0^2=(N-2)^2/N^2$, $w_\ell^2=4/N^2$, and
\begin{equation}\label{}
  \max E_\textsc{ii}=  \frac{N^3(3-N)}{2(N^2-1)(N+1)}.
\end{equation}

For $h_n^2$ on sites $n$ and $n+1$,
there are also two cases.
For case (I), the energy is
\begin{align}\label{}
  F_\textsc{i}&:=(\frac{N^2}{N^2-1})^2(\frac{3(N^2-1)}{4N^2}+ \sum_{sc} \text{tr}(t^ct^s V^\dagger t^st^c V)/N ),
\end{align}
and it takes the same value as that on ground states.
For case (II), the energy is
\begin{align}\label{}
  F_\textsc{ii}&:=(\frac{N^2}{N^2-1})^2(\frac{3(N^2-1)}{4N^2}+ \sum_{sc} \text{tr}(t^ct^s W^\dagger t^st^c W)/N).
\end{align}
We find, contrary to $h$,
the value of $h^2$ gets smaller when $f$ varies,
and the minimum is at $f=1/2$.

The terms $E_\textsc{i}$ and $F_\textsc{i}$ have the same value as those on ground state,
so the energy cost $E_c$ comes from $E_\textsc{ii}$ and $F_\textsc{ii}$
\begin{equation}\label{}
  E_c:=E_\textsc{ii}+\frac{2}{3N}F_\textsc{ii}+\frac{N}{3}.
\end{equation}
One crucial property is that the decrease amount by $\frac{2}{3N}F_\textsc{ii}$
is smaller than the increase amount by $E_\textsc{ii}$,
so the net effect is an increase of energy.
At $f=\frac{1}{2}$ we find the energy cost is
\begin{equation}\label{}
  E_c=\frac{4N(N^2+1)}{3(N^2-1)(N+1)}-\frac{N^3}{2(N^2-1)^2}.
\end{equation}
It is a monotone increasing function of $N$,
and in the large-$N$ limit, this goes to $4/3$.
We see that, the energy cost is independent of the system size
since the twist is a kind of perturbation,
and only one energy term contribute to the energy barrier.
This does not imply that a phase error is easy to introduce by the environment,
since the phase gate has to be done in such a particular way by the twist.

\subsection{Twist on excitations}
\label{sec:twadj}

Now we turn to see how the twist act on some excited states.
First, we analyze the effect of twist on a generic adjointor state
$|\mathbf{A}(\alpha)\rangle=\sum_n a(\alpha,n) |\mathbf{A}(\alpha,n)\rangle$
with coefficients $\{a(\alpha,n)\}$.
We consider the case $|a(\alpha,n)|^2\in O(1/L)$ $\forall n$.
We find
\begin{equation}\label{}
\langle \mathbf{A}(\alpha)|U_\textsc{tw} |\mathbf{A}(\alpha)\rangle=
  \frac{1}{L}\sum_n\langle \mathbf{A}(\alpha,n)|U_\textsc{tw} |\mathbf{A}(\alpha,n)\rangle+
  \frac{1}{L}\sum_{n\neq m} \langle \mathbf{A}(\alpha,n)|U_\textsc{tw} |\mathbf{A}(\alpha,m)\rangle.
\end{equation}
The terms in $\sum_{n\neq m}$ are zero as it is proportional to
$\langle \lambda| (V_n^\dagger t^\alpha V_n \otimes\mathds{1}) |\lambda\rangle
\langle \lambda|(\mathds{1} \otimes t^{\alpha*} )|\lambda\rangle$,
for $|\lambda\rangle$ as the eigenvector of $\mathcal{G}$,
and in the large-$L$ limit
$\langle \lambda| (\mathds{1} \otimes t^{\alpha*} )|\lambda\rangle=0$.
With
\begin{equation}\label{}
  \langle \mathbf{A}(\alpha,n)|U_\textsc{tw} |\mathbf{A}(\alpha,n)\rangle=
  2N e^{i\ell L/N} \langle \lambda|(V_n^\dagger t^\alpha V_{n}\otimes t^{\alpha*} )|\lambda\rangle,
\end{equation}
we find in the large-$L$ limit
\begin{equation}\label{}
  \langle \mathbf{A}(\alpha)|U_\textsc{tw} |\mathbf{A}(\alpha)\rangle
  \approx e^{i2\pi/N} \frac{2}{L}\sum_n \text{tr} V_n^\dagger t^\alpha V_{n} t^\alpha
  \approx e^{i2\pi/N}.
\end{equation}
Note for the adjointor from the other ground state, this value is $e^{-i2\pi/N}$.
This shows that the phase factor $e^{\pm i2\pi/N}$ is robust against single adjointor excitation.
This is also expected from Remark~\ref{remark:homotopy}, namely,
the twist phase can be obtained by acting on a segment without the adjointor excitation.

For the twist on dimer states,
we expect a trivial phase as a dimer state contains equal number of bonds from
$|\mathbf{L}\rangle$ and $|\mathbf{R}\rangle$,
Indeed we find
\begin{equation}\label{}
  \langle\mathbf{D}|U_\textsc{tw}|\mathbf{D}\rangle
  =(\text{tr}\mathcal{V}^\dagger \mathcal{M}\mathcal{V} \mathcal{M} )^{L/2}
  /(\text{tr}\mathcal{M}^2 )^{L/2}
  =(1-\frac{4}{N+1}\sin^2\frac{\ell}{2})^{L/2}=1,
\end{equation}
in the large-$L$ limit.
This can also be understood using the fermion operator method.
In this method, the dimer state can be viewed as a combination of half $|\mathbf{L}\rangle$ and half $|\mathbf{R}\rangle$.
The phase factor will be
\begin{equation}\label{}
(\frac{N-1+e^{i\ell}}{N})^{L/2}(\frac{N-1+e^{-i\ell}}{N})^{L/2}
=(1-\frac{4(N-1)}{N^2}\sin^2\frac{\ell}{2})^{L/2}=1.
\end{equation}
This means that for each double bond the phase $e^{-i\ell}$
will cancel the phase $e^{i\ell}$ in the order $O(1/L)$.

For domain wall and multiple adjointor excitations, however, the phase factor is not robust.
For instance, for a state with domains,
the state can be that with roughly half of the chain from $|\mathbf{L}\rangle$,
and half from $|\mathbf{R}\rangle$,
and also with a soliton and antisoliton separating them.
After a twist, the phase accumulated from $|\mathbf{L}\rangle$ and $|\mathbf{R}\rangle$
will cancel each other,
similar with the case of dimer states.

\renewcommand{\theequation}{SC\arabic{equation}}
\renewcommand{\thefigure}{SC\arabic{figure}}
\renewcommand{\thetable}{SC\arabic{table}}

\section{Simulation of error correction}
\label{sec:simlag}

In this section we present simulations of errors on VBSQ.
We first consider a discrete set of error in section~\ref{subsec:diserror},
and then more general errors in section~\ref{sec:gerror},
for which we analyze the approximate error correction behavior for dilute errors.

\subsection{Discrete set of error}
\label{subsec:diserror}

The Heisenberg-Weyl operators $X^jZ^k$, known as generalized Pauli operators,
are defined by
\begin{equation}\label{}
  X^j:=  \sum_{\alpha=0}^{d-1} |\alpha\rangle\langle \alpha+j|, \;
  Z^k:=  \sum_{\alpha=0}^{d-1} \omega^{k\alpha} |\alpha\rangle\langle \alpha|,
\end{equation}
for $\omega=e^{i2\pi/d}$, $j,k\in \mathbb{Z}_d$,
and $X^j Z^k=\omega^{jk} Z^k X^j$.
Note in our case $d=N$.
The basis index $\alpha$ can be interpreted as the flavor index
of the virtual valence bond.
We will denote a Pauli operator as $P^{jk}=X^jZ^k$,
and the adjoint version of $X$, $Z$ and $P^{jk}$
as $\mathscr{X}$, $\mathscr{Z}$, and $\mathscr{P}^{jk}$, respectively.
The error set $\mathcal{E}:=\{\Pi,\mathscr{P}^{jk}\}$,
for the bit-flip operator $\Pi$ from the parity symmetry,
is correctable and allows code distance $L$.
The set $\mathcal{E}$ is noncommutative as
\begin{equation}\label{}
\Pi \mathscr{X} \Pi = \mathscr{X},\; \Pi \mathscr{Z} \Pi = \mathscr{Z}^\dagger.
\end{equation}

\begin{figure}[t!]
  \centering
  \includegraphics[width=.4\textwidth]{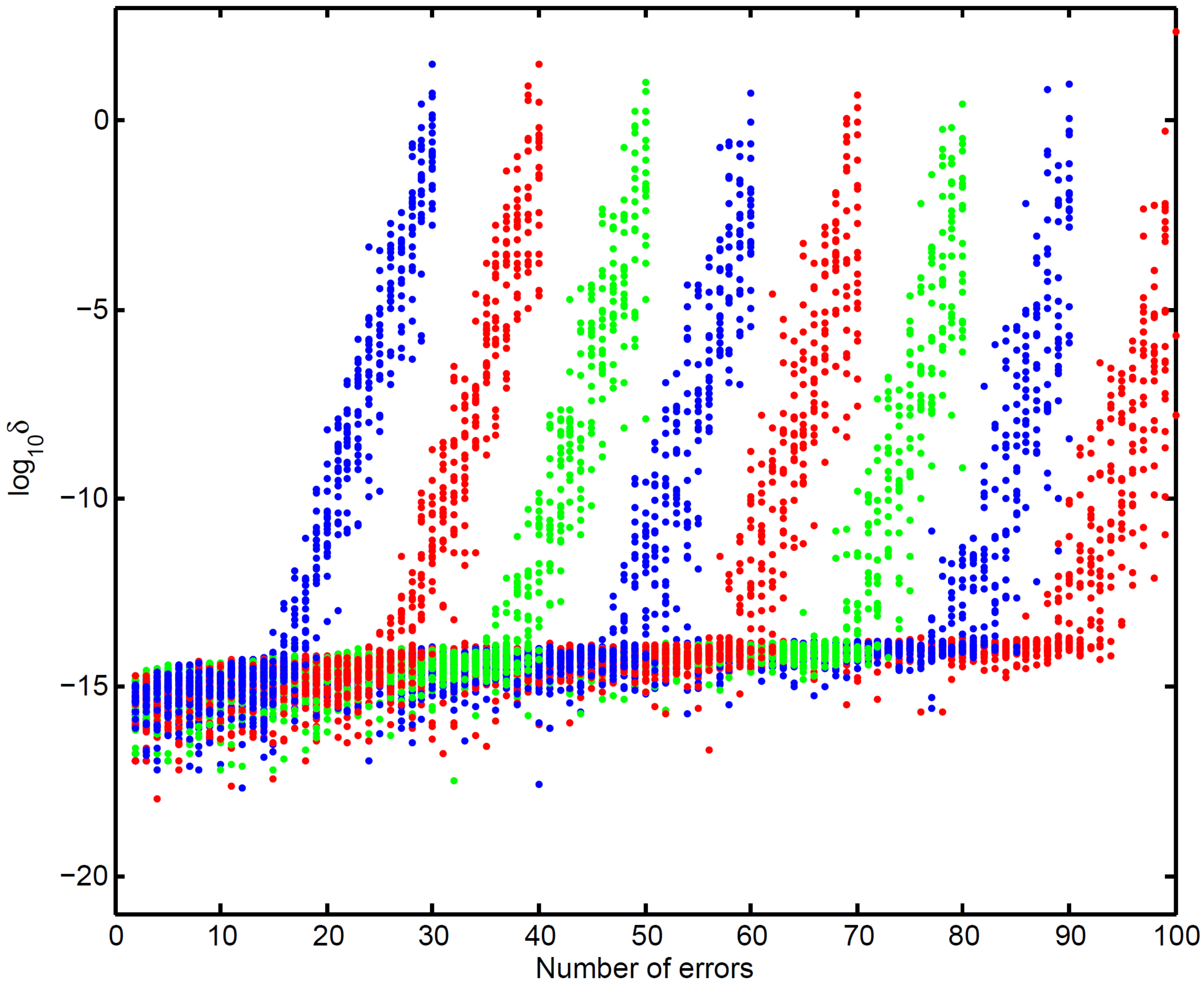}
  \includegraphics[width=.405\textwidth]{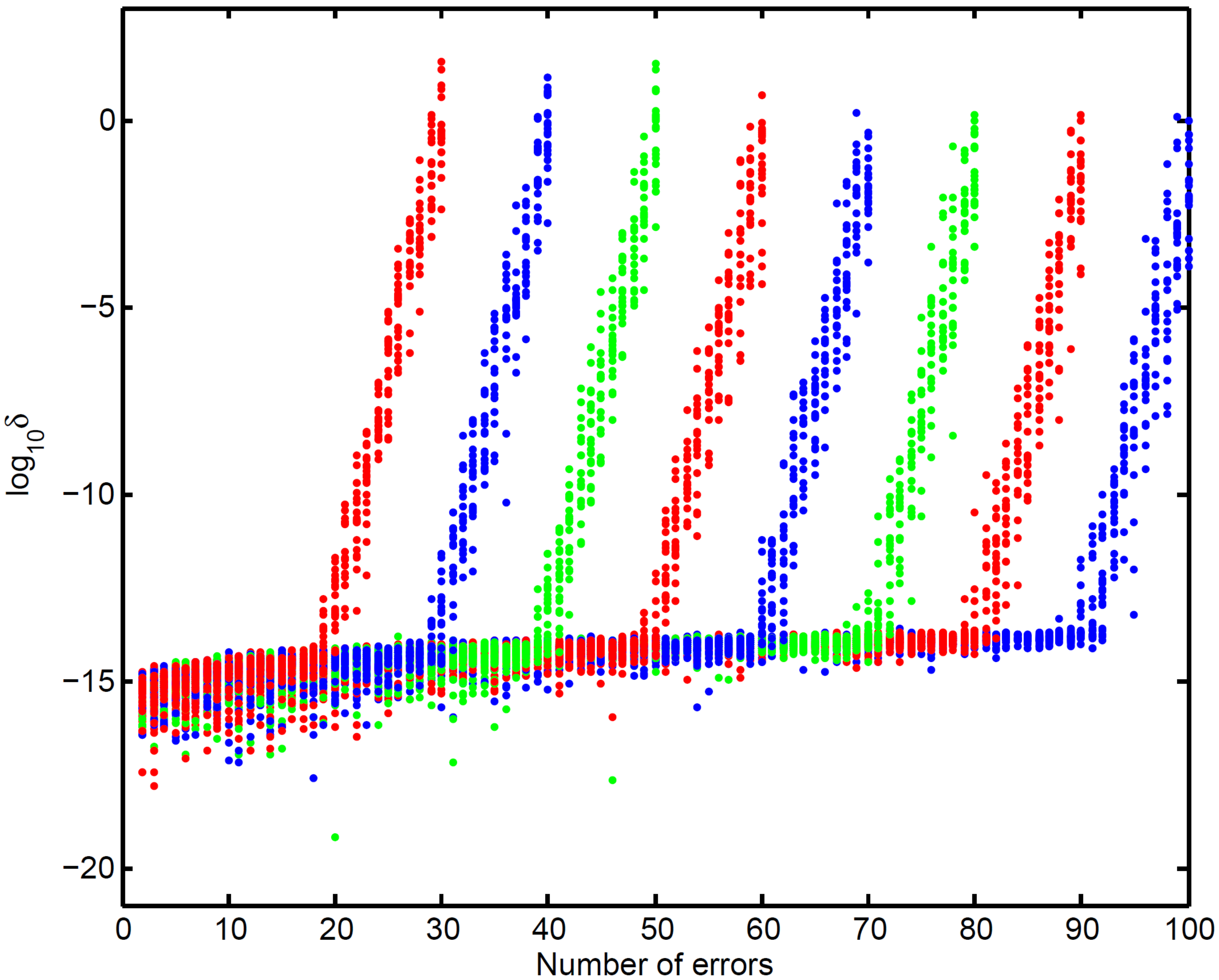}
  \caption{Simulation of the deviation $\log_{10}\delta$ of $\mathbf{U}(n)$ from identity
  for $SU(4)$ (left) and $SU(5)$ (right) cases.
  The system size increases (from left to right) from $L=60$ to $200$ by an increase of amount $20$.
  For a fixed number of errors $n$,
  we chose 20 random errors at $n$ random positions.
  }\label{fig:su45}
\end{figure}

We performed numerical simulations of the error-correction properties of the set $\mathcal{E}$
for $SU(3)$, $SU(4)$, and $SU(5)$ cases.
The $SU(3)$ case has been shown in the main text,
while here we present the $SU(4)$ and $SU(5)$ cases,
see Fig.~\ref{fig:su45}.
Similar with the $SU(3)$ case,
we can see that an increase of the system size by 20
corresponds to an increase of the critical number of errors by 10.
This means that the critical number of errors is $L/2$,
and then the code distance for $\mathcal{E}$ is $L$.
In addition, some details are as follows.
\begin{itemize}
  \item $SU(3)$.
  In Pauli basis,
  the Pauli error $\mathscr{X}=\text{diag}(1,1,\omega^2,\omega^2,\omega,\omega,\omega^2,\omega)$,
  $\mathscr{Z}=\text{diag}(\omega,\omega^2,\omega,\omega^2,\omega,\omega^2,1,1)$, and
  the bit-flip operator $\Pi$ takes the form
  $\Pi=\text{diag}(\sigma_x,A,A^\dagger,\mathds{1})$ for Pauli $\sigma_x$ and $A=[0,\omega;\omega^2,0]$
  in the basis $\{X,X^2,XZ,X^2Z,XZ^2,X^2Z^2,Z,Z^2\}$.

  First, for the error $\Pi$,
  let $\Pi(r) $ denote $r$ errors $\Pi$ at random positions.
  For $\mathbf{G}\in\{\mathbf{L},\mathbf{R}\}$,
  we find the following general scalings
  \begin{align}\label{}
  \langle \mathbf{G}|\Pi(r) |\mathbf{G}\rangle&=\left\{
                \begin{array}{ll}
                  (\frac{1}{N+1})^{3L/2-r},& (r> \frac{3}{4}L ),\\
                  (\frac{1}{N+1})^r, &        (r< \frac{3}{4}L ),\;
                \end{array}
              \right. \\
    |\langle \mathbf{L}|\Pi(r) |\mathbf{R}\rangle|&=\left\{
                \begin{array}{ll}
                  (\frac{1}{N+1})^{L/2+r}, & (r< \frac{1}{4}L ),\\
                  (\frac{1}{N+1})^{L-r},   &        (r> \frac{1}{4}L ).
                \end{array}
              \right.
  \end{align}
  Also $|\text{Im}\langle \mathbf{L}|\Pi^r |\mathbf{R}\rangle|$ $\ll$
  $|\text{Re}\langle \mathbf{L}|\Pi^r |\mathbf{R}\rangle|$.

  For Pauli errors, let $\mathscr{P}(r)$ denote $r$ random Pauli errors at random positions.
  We find
  \begin{align}\label{}
  \langle \mathbf{G}|\mathscr{P}(r) |\mathbf{G}\rangle&=\left\{
                \begin{array}{ll}
                  (\frac{1}{N^2-1})^{2L \eta  -r}, & (r> \eta L), \eta=\log_{4}3,\\
                  (\frac{1}{N^2-1})^r, & (r< \eta L),
                \end{array}
              \right. \\
    |\langle \mathbf{L}|\mathscr{P}(r) |\mathbf{R}\rangle|&=\left\{
                \begin{array}{ll}
                  (\frac{1}{N+1})^{L/2+r}, & (r< L (\eta -1/2) ),\\
                  (\frac{1}{N})^{L}, & (r>L (\eta -1/2) ).
                \end{array}
              \right.
  \end{align}
  In this case, we find
  $|\text{Im}\langle \mathbf{L}|\mathscr{P}(r) |\mathbf{R}\rangle|$ $\approx$
  $|\text{Re}\langle \mathbf{L}|\mathscr{P}(r) |\mathbf{R}\rangle|$.
\item $SU(4)$.
  In Pauli basis,
  the bit-flip operator $\Pi=\text{diag}(\sigma_x,A,B,A^\dagger,C,C,\mathds{1}_3)$
  for Pauli $\sigma_x$ and
  $A=[0,\omega;\omega^3,0]$,
  $B=[0,\omega^2;\omega^2,0]$,
  $C=[1,0;0,\omega^2]$
  in the basis $\{X,X^3$, $XZ,X^3Z$, $XZ^2,X^3Z^2$, $XZ^3$,
  $X^3Z^3,X^2,X^2Z,X^2Z^2,X^2Z^3,Z,Z^2,Z^3\}$.
  We find
  \begin{align}\label{}
     |\langle \mathbf{L}|\Pi(r) |\mathbf{R}\rangle|&=\left\{
                \begin{array}{ll}
                  (\frac{1}{N+1})^{\eta L+r}, & (r< \frac{1}{2}(1-\eta)L ), \eta=0.68,\\
                  (\frac{1}{N+1})^{L-r}, & (r> \frac{1}{2}(1-\eta) L ),
                \end{array}
              \right.
  \end{align}
  while there is no concise behavior for Pauli errors.
  \item $SU(5)$. For this case, the formula of $\Pi$ and $\mathscr{P}$ are a little complicated,
  so it is not necessary to report them.
  For the general scaling, we find
   \begin{align}\label{}
    |\langle \mathbf{L}|\Pi(r) |\mathbf{R}\rangle|&=\left\{
                \begin{array}{ll}
                  (\frac{1}{N+1})^{\eta L+r}, & (r< \frac{1}{2}(1-\eta) L ), \eta=0.77,\\
                  (\frac{1}{N+1})^{L-r}, &  (r>  \frac{1}{2}(1-\eta) L ).
                \end{array}
              \right.
  \end{align}
\end{itemize}

In addition, we also simulated the set
$\mathcal{E}':=\{E_\ell|E_\ell=\Pi^x \mathscr{P}^{jk},x=0,1\}$.
First, we find the error set $\mathcal{E}'$ is 1-correctable as
the error correction condition is satisfied
\begin{equation}\label{}
  \langle \mathbf{L}|E_\ell E_\gamma |\mathbf{L}\rangle=
  \langle \mathbf{R}|E_\ell E_\gamma |\mathbf{R}\rangle,\;
  \langle \mathbf{L}|E_\ell E_\gamma |\mathbf{R}\rangle=0.
\end{equation}
However, our simulation for $SU(3)$, $SU(4)$, and $SU(5)$ cases shows that
this set $\mathcal{E}'$ does not allow bigger code distance.
For instance, the set $\{\Pi \mathscr{X}, \Pi \mathscr{X}\mathscr{Z} \}$
does not allow code distance $L$.

\subsection{More general errors}
\label{sec:gerror}

For a unitary error $U\in SU(N^2-1)$ acting on the physical space, we find
\begin{align}\label{}
  \langle \mathbf{L}|U|\mathbf{L}\rangle=
  \langle \mathbf{R}|U|\mathbf{R}\rangle=\frac{\text{tr} U}{N^2-1},
  \langle \mathbf{R}|U|\mathbf{L}\rangle=0.
\end{align}
This means unitary errors $U\in SU(N^2-1)$ are detectable.
However, those errors are not correctable,
demonstrated by our simulation below in Table~\ref{tab:sim}.

For errors in the symmetry,
a unitary operator $V\in SU(N)$ on the virtual space
can be written as a linear combination of Pauli errors
$V=v_0 \mathds{1}+\sum_i v_i P_i$.
For error detection, we find
\begin{align}\label{}
  \langle \mathbf{L}|\mathscr{V}|\mathbf{L}\rangle
  =\langle \mathbf{R}|\mathscr{V}|\mathbf{R}\rangle
  =\frac{|v_0|^2N^2-1}{N^2-1},\;
  \langle \mathbf{L}|\mathscr{V}|\mathbf{R}\rangle=0,
\end{align}
which means errors in $SU(N)$ is detectable.
Note that given $V\in SU(N)$ on the virtual space
its adjoint rep $\mathscr{V}$ has matrix elements
  $(\mathscr{V})_{ij}=\text{tr}(P_i^\dagger V P_j V^\dagger)/N$
in Pauli basis $\{P_i\}$.
Note here $\{P_i\}$ is the generalized Pauli matrix basis
for the virtual space of dimension $N$.
Furthermore, as have been shown in the main text,
these errors are also approximately correctable
when the spatial distance between errors are big enough.
We performed numerical simulations to reveal these features.
Table~\ref{tab:sim} shows the simulation of error detection and correction for the $SU(3)$ case.
We see that single random unitary in $SU(8)$ is detectable while not correctable.
Random unitary in $SU(3)$ is not correctable if their spatial distance is small.

\begin{table}[b!]
  \centering
  \begin{tabular}{||c||c|c|c|}
    \hline \hline
             & $SU(8)$ & $SU(8)\times SU(8)$ & $SU(3)\times SU(3)$ \\ \hline \hline
    $u_{LL}$ & tr$U$/8 & $a$, $10^{-2}$        & $a$, $10^{-2}$ \\ \hline
    $u_{RR}$ & tr$U$/8 & $b$, $10^{-2}$        & $a^*$, $10^{-2}$ \\ \hline
    $u_{LR}$ & 0       & $c$, $10^{-2}$        & 0 \\ \hline
    $u_{RL}$ & 0       & $d$, $10^{-2}$        & 0 \\
    \hline
  \end{tabular}
  \caption{Simulation of error detection and correction for $SU(3)$ VBC.
  The size of matrix elements $u_{ij}\in \mathbb{C}$ $(i,j=L,R)$
  is an estimation based on multiple simulations
  for various errors:
  (1st column) random unitary in $SU(8)$,
  (2nd column) two random unitaries in $SU(8)$ with arbitrary spatial distance,
  (3rd column) two random unitaries in $SU(3)$ for nearest-neighbor sites.
  }
  \label{tab:sim}
\end{table}

\begin{figure}[t!]
  \centering
  \includegraphics[width=.5\textwidth]{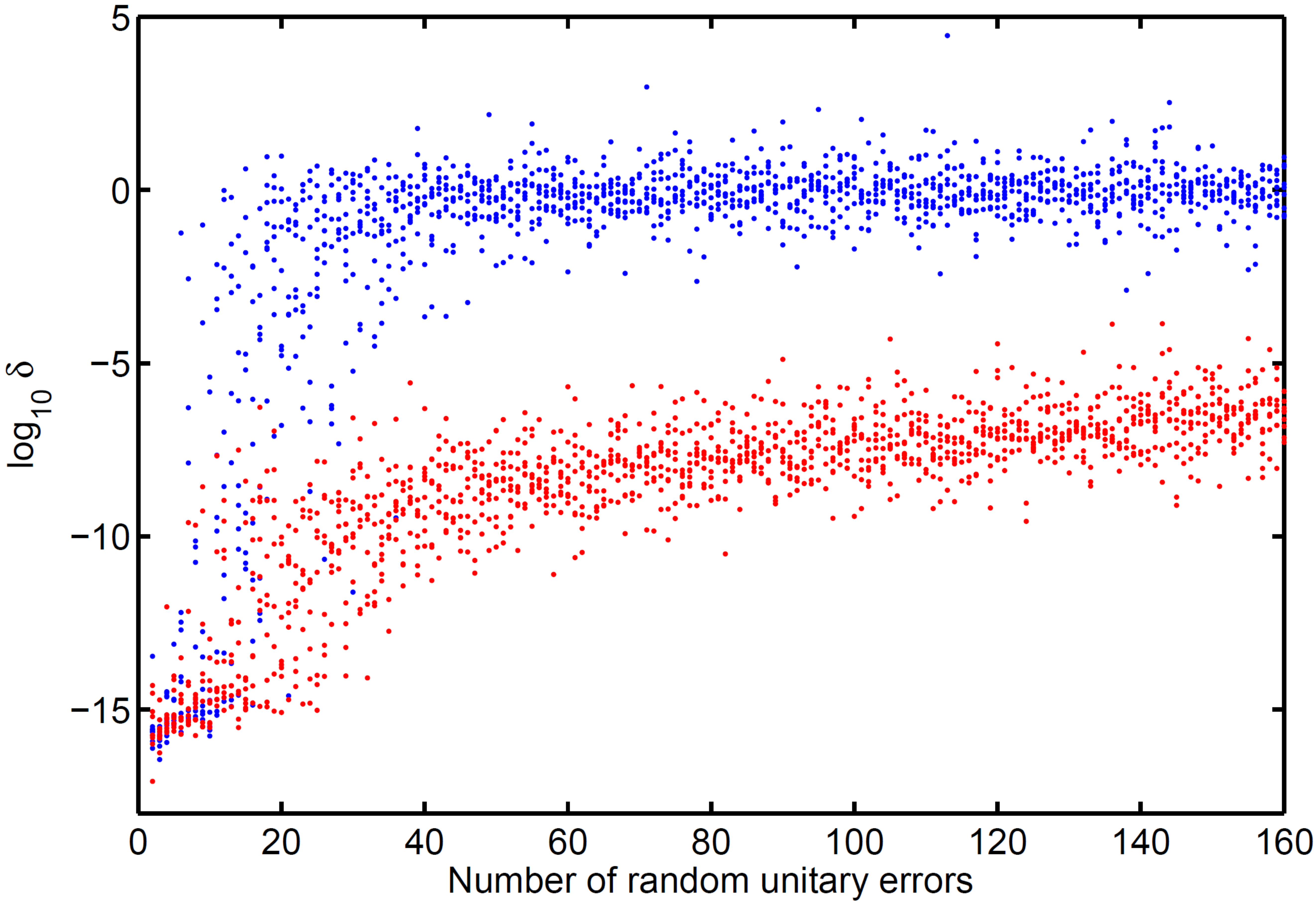}
  \caption{Simulation of violation of quantum error correction condition
  for many random unitary errors for the $SU(3)$ case.
  The lower bound for the distance between two random errors is 0 (blue dots)
  or 10 (red dots).
  The system size is $L=2000$, and for each number $n$ we sample 10 random errors.}\label{fig:ranu}
\end{figure}

For many random unitary errors from the symmetry,
we expect that the code is more robust against dilute errors.
From Fig.~\ref{fig:ranu} we see that the effective gate deviates from identity quickly
if errors are dense,
while for dilute errors, e.g.,
with a lower bound of their distance $r=10$ shown in the figure,
the effective gate remains close to identity for a certain growing number of errors.
We find the effective gate is a phase rotation
$p\text{diag}( e^{i\theta}, e^{-i\theta})$ as
the off-diagonal elements remain small compared with the diagonal elements.
The size $p$ decays exponentially w.r.t. the number of errors,
similar with the case of Pauli errors.

For an arbitrary initial logical state
$|\psi_L\rangle=\alpha|\mathbf{L}\rangle+\beta|\mathbf{R}\rangle$,
after two random unitary errors with distance $r$
it becomes $|\psi_L'\rangle$.
After cooling back to the code space, the fidelity is
\begin{equation}\label{}
  |\langle \psi'_L|\psi_L\rangle|^2=1-\mathcal{O}((N^2-1)^{-2r}).
\end{equation}
For three on-site errors with two distance $r_1$ and $r_2$,
it is not hard to see that, there are imaginary parts that
with a factor $(-\frac{1}{N^2-1})^{r_1}$, or $(-\frac{1}{N^2-1})^{r_2}$,
or $(-\frac{1}{N^2-1})^{r_1+r_2}$, and the last one is smaller.
We can see that for many on-site errors,
the dominant terms in the imaginary part come from the two-local cross terms
for all pairs of errors.
For $t$ errors, $t\in \mathbb{Z}^+$, there are $t(t-1)/2$ pairs of errors,
while there are $t-1$ neighboring pairs of errors,
and the errors from neighboring pairs will dominant.
Note here $t$ should not be confused with Gell-Mann matrices.
We can let $r$ be a constant, e.g., $r=10$,
and then we can see that $t:=\lfloor (L-1)/(r+1)\rfloor$ errors can be approximately corrected,
so the code distance for approximate error correction is
\begin{equation}\label{eq:codedis}
d_\text{app}:=2t+1=2\lfloor (L-1)/(r+1)\rfloor+1,
\end{equation}
so it is linear with $L$.
This behavior can be put into the framework of
\emph{approximate} code~\cite{CGS05,BO10,CL14} for
arbitrary on-site unitary errors in the symmetry.
The distance $r$ between errors is a parameter in the approximation.
For nontrivial errors $\{E_i\}$,
each of which can be a product of $n$ errors
at $n$ different sites $r_1,\dots, r_n$,
the approximate error-correction condition is defined as
\begin{equation}\label{eq:aerc}
  \langle\psi_a| E_i^\dagger E_j |\psi_b\rangle= C_{ija}\delta_{ab},
  |C_{ija}-C_{ijb}|\in \mathcal{O}(\epsilon),\;
  \forall |\psi_{a(b)}\rangle\in \mathcal{C},
\end{equation}
for $\epsilon:= (N^2-1)^{-r}$, $r$ as the minimal distance of sites from $E_i$ and $E_j$.

Finally, we emphasize that for exact error correction
concatenation with other codes can be employed to deal with phase errors,
as discussed in the main text.
For concatenation many VBSQ rings are needed,
and the error detection of any single ring for phase error will benefit
the concatenated codes.
By keeping the system at a low temperature,
there will be a pattern of dilute errors
and phase errors on single VBSQ ring can be suppressed.

\subsection{Energy penalty syndrome}
\label{sec:errorenergy}

Local errors can be detected by measuring the energy of each term $H_n$.
Here we compute the energy penalty of generalized Pauli errors.
A Pauli error $\mathscr{P}_n$ on site $n$ will affect
the two-body energy terms $H_{n-1}$ and $H_n$.
We find that the energy penalty is the same for the two ground states,
so we let $|\mathbf{G}\rangle\in\{|\mathbf{L}\rangle,|\mathbf{R}\rangle\}$.
We find
\begin{equation}\label{eq:paulienergy}
\langle \mathbf{G}|\mathscr{P}^\dagger_n h_n \mathscr{P}_n|\mathbf{G}\rangle
=\frac{N^3}{2(N^2-1)^2}.
\end{equation}
This is the same with the energy penalty of an adjointor.
Also note that when
two Pauli errors $\mathscr{P}_1$ and $\mathscr{P}_2$ act on two neighboring sites $n$ and $n+1$,
the effective error is $\mathscr{P}:=\mathscr{P}_2^\dagger \mathscr{P}_1$ on site $n$,
so the energy penalty of $h_n$ or $h_{n-1}$ is still $\frac{N^3}{2(N^2-1)^2}$.
If $\mathscr{P}_1=\mathscr{P}_2$, then there is no energy change for the term $h$ on the two sites,
while the energy penalty will be shown on the two neighboring terms.
With the same method we find
\begin{equation}\label{eq:paulienergy2}
\langle \mathbf{G}|\mathscr{P}^\dagger_n h_n^2 \mathscr{P}_n |\mathbf{G}\rangle
=\frac{N^2(3N^2-2)}{4(N^2-1)^2}.
\end{equation}
This is also the same with the energy penalty of an adjointor.
This then means that a Pauli error causes about twice the energy of an adjointor.

Now we consider random unitary error $V\in SU(N)$.
With $V=c_0 \mathds{1}+\sum_\ell c_\ell P_\ell$,
 the energy penalty is
\begin{align}\label{eq:verrorlap1}
  \langle \mathbf{G}|\mathscr{V}^\dagger_n h_n \mathscr{V}_n |\mathbf{G}\rangle
=  -|c_0|^2\frac{N^3}{2(N^2-1)}+\frac{N^3}{2(N^2-1)^2} \sum_\ell |c_\ell|^2.
\end{align}
We can see the first term is the energy for no error,
and the second term is the energy for those Pauli errors.
When there are two errors $\mathscr{V}_1$ and $\mathscr{V}_2$ acting on nearest sites,
then the effective error is $\mathscr{V}=\mathscr{V}_1^\dagger \mathscr{V}_2$.
Similarly, for $h_n^2$ the energy is
\begin{align}\label{eq:verrorlap2}
  \langle \mathbf{G}|\mathscr{V}^\dagger_n h_n^2 \mathscr{V}_n |\mathbf{G}\rangle
=\frac{3N^2}{4(N^2-1)}  +\frac{N^2}{4} |c_0|^2+
\frac{N^2}{4(N^2-1)^2} \sum_\ell |c_\ell|^2.
\end{align}
For $c_0=1$ this reduces to the energy on ground states,
and for Pauli errors this reduces to the energy of Pauli errors.

\renewcommand{\theequation}{SD\arabic{equation}}
\renewcommand{\thefigure}{SD\arabic{figure}}
\renewcommand{\thetable}{SD\arabic{table}}

\section{Concatenation}
\label{sec:concat}

In the main text we have shown that
better codes can be constructed by concatenation.
By concatenation, the merits of VBSQ and stabilizer codes can be combined together.
We use the VBSQ as the inner code, or in other words,
the VBSQ serves as a physically robust qubit that has
robust (transversal, topological or SPT) logical operators.
Here we discuss the concatenation in more details.

First, we can extend the concatenation to all even $N$ cases,
with $\bar{Z}$ provided by the half-twist (Sec.~\ref{sec:ztwist}).
The purpose is to reduce the value of $N$ for gate implementation.
As the half-twist will cause a leakage of the code while without introducing any other errors,
it can be followed by a cooling on the hardware level
and error correction on the software level to remove leakage and errors.
With the half-twist, phase gate on a VBSQ can be achieved for $N=4$
and $T$ gate for $N=8$.

In this section, denote a stabilizer code as $\mathcal{C}$ and its stabilizer as $\mathcal{S}$.
A codeword $|\psi\rangle$ of $\mathcal{C}$
can be prepared by measuring its stabilizers $S\in \mathcal{S}$.
Each stabilizer can be constructed by tensor product of Pauli matrices.
For VBSQ with even $N=2r$, $r\in \mathbb{N}$,
it has logical operator $\bar{X}$, which is transversal,
and $Z$-rotation by angle $\frac{\pi}{r}$, denoted by $\bar{Z}^{1/r}$ for clarity.
The gate $\bar{Z}^{1/r}$ is from the half-twist,
and the logical $\bar{Z}$ is from $r$ runs of $\bar{Z}^{1/r}$.
Using the measurement-based scheme,
the stabilizers $S\in \mathcal{S}$ can be enacted,
and each Pauli operator in a stabilizer $S$ is protected by the VBSQ.

For $\mathcal{C}$ as a repetition code,
it requires a series of VBSQ rings in parallel.
The stabilizer is $\{\bar{X}_i\bar{X}_{i+1}\}$ for all nearest-neighbor pairs.
Using the method presented in the main text,
any stabilizer $\bar{X}\bar{X}$ can be
fault-tolerantly measured using an encoded ancilla, such as the five-qubit code.
The codeword of the repetition code is
$|0/1_\textsc{l}\rangle=\frac{1}{\sqrt{2}}(|\mathbf{U}\mathbf{U}\mathbf{U}\cdots \rangle
\pm |\mathbf{D}\mathbf{D}\mathbf{D}\cdots \rangle)$,
and the logical operator $X_\textsc{l}=\bar{X}$ for $\bar{X}$ on any ring
and logical $Z_\textsc{l}=\bar{Z}\bar{Z}\bar{Z}\cdots$.
Note we use the subscript $\textsc{l}$ for the software encoding,
in this case, the repetition code,
and the superscript bar for the VBSQ operators.
Using the VBSQ-repetition concatenation, the phase error on any ring becomes correctable
since $Z_\textsc{l}$ is transversal.
We see that VBSQ and repetition code each conquers a different error:
the VBSQ is robust against bit flip ($X$ error) as its $\bar{X}$ is transversal,
the repetition code is robust against phase flip ($Z$ error) as its $Z_\textsc{l}$ is transversal.
In addition, one may notice a similarity with Shor code~\cite{Shor95}.
Our concatenation is in a similar spirit with Shor code, yet the details are different.
Shor code is a concatenation of two classical repetition codes in different bases,
one for bit flip, and the other for phase flip.
The key feature of our scheme is the SPT order,
which leads to the topological logical phase gates, of the inner code.

For general stabilizer codes on qubits, including CSS codes, Reed-Muller codes,
and also topological codes such as the surface code,
there are constrains on transversal set of gates.
It is proved that~\cite{EK09,ZCC11,BK13,AC16}
the strongly transversal, i.e., same on each qubit,
$Z$-rotations are of angle $\frac{a\pi}{2^k}$, for integers $a,k$.
For instance, the 15-qubit quantum Reed-Muller code has
transversal $T=Z^{1/4}$ gate~\cite{BK05}.
On the other hand, the VBSQ allows transversal $Z$-rotation of angle
$\frac{a\pi}{r}$ for integers $a,r$.
As the result, the VBSQ and stabilizer codes fit when $r=2^k$,
i.e., $N=2r=2^{k+1}$.
For instance, for $N=8$ there are transversal phase gate $S$
and for $N=16$ there are transversal $T$ gate.
With the half-twist, the values are 4 and 8, respectively.
The transversal gates are constrained by the structure of the Clifford hierarchy.
For general $N$ the VBSQ provides more logical $Z$-rotations beyond the Clifford hierarchy.
For instance,
for $N=3$ the VBSQ has $Z$-rotation of angle $\frac{2\pi}{3}$,
and for $N=6$ the VBSQ has $Z$-rotation of angle $\frac{\pi}{3}$.
This demands a much deeper study of Clifford hierarchy,
which is beyond the scope of the current work.

\bibliography{ext}{}
\bibliographystyle{apsrev4-1}